\documentstyle[aps,epsfig]{revtex}
\begin{document}                                                                
%
%
\draft
\title{The Gravitomagnetic Field and Penrose Processes}

\renewcommand{\thefootnote}{\fnsymbol{footnote}}
\author{Reva Kay Williams
\footnote{                
E-mail:
revak@astro.ufl.edu.}}
\address{Department of Astronomy,
University of Florida,
Gainesville, Florida 32611.} 


\maketitle
%
%
%
%
%
\begin{abstract} 
Results from  general relativistic 
theoretical 
Monte Carlo computer simulations
of Compton scattering and $e^- e^+$ pair production processes in 
the ergosphere  of a supermassive ($\sim 10^8 M_\odot$)
rotating black hole are presented.
Particles from an accretion disk 
fall into
the ergosphere and scatter off particles that are in 
bound equatorially
and nonequatorially confined orbits.
The Penrose mechanism, in general,  allows 
rotational energy 
of a Kerr black hole  to be extracted by 
scattered particles escaping from the 
ergosphere to infinity (i.e., large distances from the black hole).  
The results of 
these model calculations  
show that a form of the Penrose mechanism is capable of 
producing the observed high energy particles 
(up to $\sim $~GeV) emitted by quasars and other active galactic nuclei 
(AGNs), without the necessity of the external electromagnetic field
of the accretion disk.  
Importantly, these model calculations show that
the Lense-Thirring effect, i.e., the dragging of 
inertial frames into rotation, caused by the 
 angular momentum of the rotating black hole, 
results in a gravitomagnetic (GM)
force being exerted on  
the scattered escaping particles.  
Inside the ergosphere, where 
this dragging is severe, in appears that the GM vector field lines
are frame dragged into the positive azimuthal direction, i.e., 
the direction of rotation of the black hole.  The resulting GM 
force acting on the Penrose scattered particles  produces 
 symmetrical and asymmetrical
(or one-sided) particle emissions in the
polar direction, consistent with the astrophysical
 jets observed in radio strong AGNs.
Note, these
Penrose processes can apply to any size rotating black hole. 
\end{abstract}

\narrowtext
\mediumtext
\widetext
%
%
%
%
\section{Introduction}
\label{sec:1}

In 1918, it was discovered in
the equations of general relativity by Thirring and Lense [1] 
that the angular momentum of a massive body causes inertial frames to 
be
dragged around in the direction that the mass is rotating.  
In 1969, Roger Penrose [2] noted that, inside the ergosphere of
a rotating black hole,
this frame dragging is so extreme that it can provide  
a way to extract rotational 
energy (we now call this the Penrose mechanism).  
In this paper, I present the results of the first {\it complete},
fully general relativistic, four-dimensional, theoretical
and numerical model calculation using the Penrose mechanism
to extract high energy electrons and photons 
from a rotating black hole. 
Penrose scattering
processes in the ergosphere allow high energy particles to escape to 
infinity
(far away from the black hole) with a portion of the 
rotational energy of the Kerr
black hole (KBH), and orbital energy-momentum produced by 
its strong gravitational field.
This  model calculation not only confirms the
results of energy extracted from 
Penrose Compton scattering processes
given in [3,4], however, extracting higher energies when nonequatorially 
confined targets are introduced,
but this model gives space momenta of individual scattered particles,
thus, in this sense,  making it a complete calculation.  
Importantly,
the Penrose pair production, 
by photons in bound unstable {\it trapped}  
 orbits with
photons on
radially infalling geodesics,
at the {\it photon orbit}, allow electron-positron ($e^-e^+$)
pairs to escape with 
energies
as high as $\sim 4$~GeV (implying a Lorentz factor $\gamma
_e\sim 10^4$), or higher depending on properties of the accretion
disk.   
Up until the investigation
reported here and the related paper of [5],
such Penrose pair production (PPP) at the photon
orbit had yet to be proposed.  The reason for this is probably
because the trajectories for massless and material 
particles, in nonequatorially confined
orbits, i.e., orbits not confined to the equatorial plane of the KBH, 
had to be solved first to obtain analytical
expressions for the conserved energy $E$  and conserved azimuthal 
angular momentum $L$, as measured by an observer at infinity; 
this is done in [5], and the results are
presented again here in Section~\ref{sec:2a}.

The $\it classical$ Penrose
process utilizes the existence of retrograde particle orbits (with
respect to the rotation of the KBH) in the ergosphere, for which the
energy, as would be measured
by an observer at infinity, is negative
[2,3].
Such orbits do not come out
to infinity. 
However, it is
possible for a particle, say $p_1$, that has fallen inwardly from
infinity into the ergosphere to scatter off another particle, say
$p_2$, initially in a direct orbit inside the ergosphere.
If the orbit of the scattering particle $p_2$ changes into a
retrograde orbit (of negative energy), then the scattered
particle $p_1$ can escape to infinity with more mass-energy than
the sum of the initial energies of $p_1$ and $p_2$. Since the  orbit
of the initially bound particle $p_2$ is dependent on the angular
momentum of the KBH and the curvature of spacetime defined by the
mass  of the  hole, 
when $p_2$ gives up energy to the escaping particle $p_1$ and falls
into the event horizon, the KBH loses energy in the form of
rotational energy.  

One plausible class of Penrose processes occurs when particles
already inside the ergosphere (say particles of an accretion
disk) undergo {\it local relativistic scatterings} [6,7]. 
Such types are
considered in this investigation. If one of the scattering
particles is initially in a bound orbit, then it is possible for the
other initially unbound scattered particle, or the new
particles created in the scattering process, to escape to infinity 
with
rotational energy-momentum from the KBH,
extracted either
directly (classical Penrose process) or indirectly
(quasi-Penrose process), as defined in [5].
This process allows
scattered ergospheric particles to
(1) escape to infinity with more mass-energy than they would have had
if the scattering occurred outside the ergosphere and (2) escape
to infinity with mass-energy initially trapped
by the KBH, mass-energy near the event horizon 
that had no other way of escaping, except by such processes 
as these Penrose processes;
otherwise this trapped mass-energy would
eventually be accreted into the black hole.  That is, since, in
general, nothing
can come out of the black hole (including for orbits near the event 
horizon), then
a physical process near the  hole, such as a Penrose process,
is necessary to inject a trapped or ``plunging'' orbiting
particle into 
an escaping orbit [7]. 
In other words, these so-called Penrose
processes provide a way for blueshifted mass-energy,
initially trapped by the KBH, to escape to infinity.

The model calculations discussed in this paper consist of
using the Monte Carlo
method to treat Penrose processes in the ergosphere of a KBH. The
processes investigated are
Penrose Compton scattering (PCS), $\gamma$-ray---proton 
PPP ($\gamma p \longrightarrow e^- e^+p$\/),
and $\gamma$-ray---$\gamma$-ray PPP 
($\gamma \gamma \longrightarrow e^- e^+$\/). 
These three processes are
assumed to  occur in material that has fallen into the ergospheric 
region
of a rotating black hole from a surrounding accretion disk.
The Monte Carlo method is applied to the cross section of 
each scattering event to determine values of the 
scattering
angles and final energies as measured by local observers.

In  [5], I present the detailed model 
calculations for the above
Penrose scattering processes; the final outcome 
is the four-momenta of the scattered particles as
measured by an observer at infinity.
The important properties and relations found from the model 
calculations of
[5] are described in this paper.
Now, of the three Penrose scattering processes investigated,
the most rewarding ones were the
PCS and the PPP ($\gamma \gamma \longrightarrow e^- e^+$\/): 
these two processes are specifically discussed 
in this paper. 
Unfortunately, the PPP ($\gamma p \longrightarrow e^- e^+p$\/) 
as suggested in [8,9]
did not allow any of the
scattered $e^-e^+$ pairs to escape, mainly because of the large
inward radial momentum acquired by the assumed radially infalling 
incident photon at the scattering radius, too large for the pairs to
be scattered outward by the target orbiting proton: I refer the reader
to [5]  for details concerning this scattering process.

The purpose of this paper is to discuss 
some of the astrophysically
interesting features found by looking carefully
 at the emission spectra of the 
Penrose scattered particles.  
Two of the most important features of
the scattered particles are the following: (1)~large 
fluxes of relativistic 
$e^-e^+$ pairs
with energies as high as ${>\atop\sim}4$~GeV can escape; 
and (2) the rotation of
the KBH can cause the scattered particles to escape with asymmetrical
distributions, above and below the equatorial plane, thus, naturally 
aiding in the production of
one-sided jets of relativistic particles.  
Feature (2) is a general relativistic effect caused by the 
gravitomagnetic field [10] (i.e., the gravitational analog
 or
resemblance
of a magnetic field), due to the  angular momentum 
of the rotating black hole. 
This gravitomagnetic (GM) field, as described in 
Section~\ref{sec:2d},
 causes the dragging of
inertial frames, while at the same time produces a 
force which acts on the momentum vector of a 
particle in its field, and 
which, for small velocities ($\ll c$), is linear in 
the particle's velocity.
Both of the above features 
are important ingredients in
what would be needed 
 to explain the 
observed spectra of quasars and other AGNs.  
Note, feature (2) implies 
that,
near the event horizon the expected  reflection symmetry of the Kerr
metric,
above and below the equatorial plane,
appears to be broken due to the GM force
field.   

It has been suggested [11] that
the one-sided or asymmetrical jets of radio strong
quasars and radio galaxies
may be intrinsically related to the energy source. 
The reason for this suggestion is that 
many of the observed one-sided jet distributions of AGNs do not
conform with that predicted if the one-sidedness is due solely to 
relativistic beaming---of the emission axis of the jet 
into a direction near the line of sight
of the observer---wherein Doppler boosting of the radiation
 causes the approaching
jet to be apparently brighter than the receding jet.  Yet, up until 
the
investigation reported here, no  model has been proposed for 
the
energy source that would exhibit such asymmetrical properties.   
Herein, however, from these Penrose scattering processes [5], 
I  present 
evidence that the rotating black hole core energy source 
may be  producing two of the important observed features 
of AGNs: 
 the large energy fluxes of $e^-e^+$
pairs inferred to emerge from their cores, 
and  asymmetry in their jet
and counter jet distributions.
Moreover, findings by Dennett-Thorpe et~al. [12],
after studying the asymmetry of jets, lobe length, and spectral 
index in a sample of quasars, with well defined jets, led them  
to conclude that the simplest models of relativistic beaming fail to   
explain the ``present observations''; and that only 
if the intrinsic spectrum is curved (i.e., with the spectral index 
increasing with frequency), such that the approaching lobe is seen 
at a significant lower frequency, will asymmetry of the right sense
be produced.     
This somewhat supports the findings presented here,
that  asymmetry in the jet and counter jet is intrinsic to 
the energy source, suggesting that, relativistic beaming 
may  just serve as an important enhancement 
mechanism.    
Further, recent observations of 
so-called microquasars in our Galaxy show some asymmetry
in the jet and counter jet distributions [13];
and not  all of the observed 
asymmetry can be explained by relativistic beaming, suggesting 
that some asymmetry is intrinsic [14].

The remaining
structure of this paper is as follows:
In Section~\ref{sec:2}, I present general formalisms
and descriptions 
of the Kerr metric, particle orbits, 
the Penrose Compton scattering (PCS), the Penrose 
$\gamma$-ray---$\gamma$-ray pair
production [PPP ($\gamma\gamma\longrightarrow e^-e+$)], 
and the gravitomagnetic field (GM).  
Presented in Section~\ref{sec:3} is a 
discussion of the gravitomagnetic field as it relates 
specifically to
PCS and PPP ($\gamma\gamma\longrightarrow e^-e+$).    
Finally, in Section~\ref{sec:4}, I conclude with an overview,
including general discussions of how well these model results
agree with astronomical observations, and the role of an 
external accretion 
disk magnetic field.
Also,  suggestions are made for further investigations.        
%

%
\section{Formalism}
\label{sec:2}

\subsection{The Kerr Metric and Particle Orbits}
\label{sec:2a}

For completeness, review, and to be referred to in the 
following sections,
the spacetime separation between events near 
the KBH is defined by the 
Kerr metric [15]. 
The Kerr metric written
in Boyer-Lindquist coordinates [16] and in
geometrical units ($G=c=1$) is
\begin{equation}
ds^2=-e^{2 \nu}\,dt^2+e^{2 \psi}(d \Phi -\omega\,dt)^2+e^{2 \mu_1}
\,dr^2+e^{2 \mu_2}\,d \Theta^2,
\label{eq:1}
\end{equation}
with
\begin{eqnarray}
e^{2 \nu}&=&{\Sigma \Delta \over A}, \label{eq:2}\\ 
e^{2 \psi}&=&\sin^2 \Theta {A \over \Sigma}, \label{eq:3}\\
e^{2 \mu_1}&=&{\Sigma \over \Delta}, \label{eq:4}\\
e^{2 \mu_2}&=&\Sigma, \label{eq:5}\\
\omega&=&{2\,Ma\,r \over A}, \label{eq:6}
\end{eqnarray}
where
\begin{equation}
A \equiv \Bigl(r^2+a^2 \Bigr)^2-a^2 \Delta \sin^2 \Theta,
\label{eq:7}
\end{equation}
$a$ is the angular momentum per unit mass parameter, $M$
is the mass of the black hole, 
and $\omega$ gives the frame dragging velocity. The quantities
$\Delta$ and $\Sigma$ are defined by
\begin{equation}
\Delta \equiv r^2-2Mr+a^2
\label{eq:8}
\end{equation}
and
\begin{equation}
\Sigma \equiv r^2+a^2 \cos^2\Theta,
\label{eq:9}
\end{equation}
respectively. In general, the parameter $a$ can have values
$0 \le (a/M) \le 1$, values which allow for the existence
of an event horizon. 
For a KBH, the event horizon
is located at $r=r_+=M+\Bigl(M^2-a^2\Bigr)^{1/2}$, the larger root
of the equation $\Delta=0$.
Here, in these Penrose processes, a canonical KBH is used, with
its limiting value ${a/M}=0.998$, as defined in [17], in the
investigation of an accretion disk around a KBH.  It is found
that, if $a/M$ is initially very close to $1$
($0.999\leq {a/M}\leq 1$), a small amount
of accretion (${\Delta M/M}\leq 0.05$) through a disk quickly spins the
hole down to a limiting state  ${a/M}\simeq 0.998$.  Conversely, if
$a/M$ is initially below this limiting value, accretion spins the hole
up toward it.  

Upon approaching a KBH from infinity, a
limit is reached 
where the angular momentum of the KBH
causes inertial frames to be dragged 
around in the direction that the
black hole rotates. 
This limit is characterized by the vanishing
of the $g_{tt}$ component in Eq.~(\ref{eq:1}):
\begin{equation} 
g_{tt}=- \left(1-{2Mr \over \Sigma} \right)=0.
\label{eq:10}
\end{equation}
More precisely, this limit is 
the larger root of $g_{tt}$,
given analytically by
\begin{equation}
r=r_o=M+\Bigl(M^2-a^2\cos^2\Theta \Bigr)^{1/2}.
\label{eq:11}
\end{equation}
The region between this limit and the event horizon is
called the ergosphere.

Inside the ergosphere, the Kerr metric in the Boyer-Lindquist
coordinate frame (BLF), i.e., frame of the observer at infinity,
does not allow an observer to be
stationary (in the sense of the observer being at rest with
$r,~\Theta,~\Phi={\rm constant}$) because of the dragging of inertial
frames.  For this reason, physical 
processes are difficult to describe
in the BLF. 
In order to examine physical processes inside the ergosphere,
Bardeen, Press, and Teukolsky [7] devised a
frame of reference called the local nonrotating frame (LNRF).
Observers in this frame rotate with the KBH in such a way that
the frame dragging effect of the rotating black hole is canceled
as much as possible.  In the LNRF, special relativity applies
since locally spacetime has Lorentz geometry.
The transformation laws, for the covariant components of a 
four-momentum, between the BLF and the LNRF tetrads are given 
in [3,5].  Note, in the LNRF, $g_{\mu\nu}=\eta_{\mu\nu}\equiv\,$
Minkowski metric components. 

However, it is better to use the BLF
when describing the general orbits of
particles (including photons) near the event horizon of the KBH.  
The BLF admits three
constants of motion as measured by an observer at infinity
[18]. In terms of the covariant components of the particle's
four-momentum
$\bigl[P_\mu=\bigl(P_r,P_\Theta,P_\Phi,P_t \bigr);\;
\bigl(\mu=r,\Theta,\Phi,t \bigr)\bigr]$ at some instant of time, the
conserved quantities are
$E=-P_t \equiv$ total energy,
$L=P_\Phi \equiv$ component of the angular momentum
                 parallel to the symmetry axis,
and
\begin{equation}
Q=P_\Theta^2 + \cos^2\Theta \biggl[a^2 \biggl(\mu_o^2-E^2 \biggr)
  +{L^2 \over \sin^2 \Theta} \biggr],
\label{eq:12}
\end{equation}
where $\mu_o$ is the rest mass
energy of the particle, which is a trivial
fourth constant of motion. ($Q$ is sometimes referred to as Carter's 
constant.) The value of $Q$ is zero for particles
whose motions are confined to the equatorial plane.  The nonzero
values of $Q$ belong to particles which are moving in the
${\pm\bf \hat e_\Theta}$ direction and/or are not 
confined to the equatorial plane.

The general expressions for the conserved 
energy $E$ and azimuthal angular
momentum  $L$
of  massless ($\mu_o=0$) 
and material particles on  orbits  not
confined  to the equatorial plane  (i.e.,with $Q>0$) in the
curved spacetime of a KBH
are given below.
For direct orbits of constant radius $r$ [5]
\begin{equation}
E=\left({r^2L^2+D+F\over G}\right)^{1/2},
\label{eq:13}
\end{equation}
and
\begin{equation}
P_\Phi= L=\left[{-J-(J^2-4IK)^{1/2}\over 2I}\right]^{1/2},
\label{eq:14}
\end{equation}
where
\begin{eqnarray*}
I&\equiv& {\tilde A^2r^4\over G^2}-{r^2\over G}(2\tilde A C
+B^2)+C^2, \\
J&\equiv& {(D+F)\over G}\Bigl[{2\tilde A^2r^2\over G}-2\tilde A C
-B^2\Bigr]+2\Delta (r^2\mu_o^2+Q)\Bigl[C-{\tilde A r^2\over G}
\Bigr], \\
K&\equiv& {\tilde A^2\over G^2}(D+F)^2-{2\tilde A \Delta\over G}
(r^2\mu_o^2+Q)(D+F)+\Delta^2(r^2\mu_o^2+Q)^2;
\end{eqnarray*}
and
\begin{eqnarray*}
\tilde A&\equiv& [(r^2+a^2)^2-a^2\Delta],\\
B&\equiv& 4Mar, \\
C&\equiv& \Delta-a^2, \\
D&\equiv& (3r^4-4Mr^3+a^2r^2)\mu_o^2, \\
F&\equiv& (r^2-a^2)Q,\\
G&\equiv& 3r^4+a^2r^2. 
\end{eqnarray*}
 Note that, when $Q=0$, Eqs.~(\ref{eq:13}) 
and~(\ref{eq:14})
reduce to the
equations for equatorially confined orbits given in [7,5].
The quantity $\sqrt{Q}$ ($\equiv P_\Theta$) versus 
the conserved energy $E\/$ 
[Eq.~(\ref{eq:13})] is plotted 
in Fig.~1(a) for the photon orbit at the radius 
$r_{\rm ph}$, and 
in Fig.~1(b) for the electron orbits at the marginally bound
and the marginally stable 
radii, $r_{\rm mb}$ and $r_{\rm ms}$,
respectively (these radii are discussed further in 
Sections~\ref{sec:2b} and~\ref{sec:2c}).
Importantly, the derivations of 
Eqs.~(\ref{eq:13}) and~(\ref{eq:14}),  derived in [5], now
allow for Penrose processes
be investigated in a practical way.  That is, the motion of
particles,  above or below the equatorial plane (in the 
${\pm\bf \hat e_\Theta}$ direction), 
can now be considered for the target particles as well as for the
scattered particles, with $E$ and $L$ 
given separately
by Eqs.~(\ref{eq:13}) and~(\ref{eq:14}), 
respectively, wherein $Q$ is 
the independent
variable for a specific black hole of mass $M$ and angular momentum 
parameter $a$, at constant
radius $r$.  

When $Q> 0$,  Eqs.~(\ref{eq:13}) and~(\ref{eq:14})
describe nonequatorially confined ``spherical-like'' orbits [19].
One of these spherical-like
orbits, near the event horizon,  consists of
a particle repeatedly passing
through the equatorial plane while
tracing out a helical belt lying on a sphere at constant radius.
 The belt width between
 the maximum and minimum latitudes that the particle
achieves, in general,  increases with increasing $Q$,
where $Q$ is given by Eq.~(\ref{eq:12}).
In these spherical-like orbits, which in the
limit of large radius goes asymptotically to a Keplerian circle,
as the orbital radius  approaches that of the event 
horizon, the dragging
of the line of nodes, in the sense of the spin of the black hole,
increases without limit.  (The nodes are points at which the orbit,
in going between negative and positive
latitudes, intersects the equatorial plane.)  Inside the
ergosphere, during the time that
a particle makes one oscillation in latitude, it will be swept through
many complete azimuthal revolutions.  Consequently, an orbit
near the event horizon will have a helix-like shape with the axis parallel
to the spin of the KBH.  In the Penrose processes investigated here,
it is assumed that a collision takes place, for a given observer, 
when the target particle with $Q>0$ passes through one of the nodes.                                          

After the scattering events, not all of the particles
escape to infinity.
There are certain conditions that a particle must satisfy 
in order for
it to be on an outgoing orbit.
A thorough discussion of the escape conditions for material and 
massless particles
is given in [5]. Generally speaking, 
these escape conditions compare
the four-momentum components $E'$, $L'$, 
and $P_\Theta^\prime$ ($\equiv\sqrt{Q'}$) 
of the scattered particle to
the values  $E$, $L$, and $Q$ of a possible turning point, 
to see if the particle is
on an escaping orbit.  
The expressions 
for $E$ and $L$ are given by Eqs.~(\ref{eq:13}) and~(\ref{eq:14}).  
Note that, for the 
nonequatorially confined ($Q>0$) spherical-like orbits, 
bound
particles can have energies $E/ \mu_o > 1$, because of an
additional escape condition that must be satisfied. 


\subsection{Penrose Compton Scattering (PCS)}
\label{sec:2b}

In the PCS process,
a photon of initial energy $E_{\rm ph}$, as 
measured by an observer at infinity (i.e., in the BLF), is assumed 
to be emitted inside the ergosphere---emitted from the innermost 
region of the accretion disk, 
and follows a null geodesic trajectory before
colliding with an equatorial or a nonequatorially confined
orbiting electron, located
at a specific 
radius of energy $E_e$ and 
azimuthal momentum $L_e$ [$= (P_e)_\Phi$] 
given by Eqs.~(\ref{eq:13}) and~(\ref{eq:14}), 
respectively. 
The motions of the incoming or incident 
photons considered in these scattering 
processes are those moving radially 
inward along the equatorial plane, as in [3].
Now, some of the photons, after being
scattered by the electrons, eventually escape to infinity. 
For the scattered photons that are allowed to escape to infinity,
an observer
there would observe
a low-energy (in most cases $< \mu_e$, where
$\mu_e$ is the electron rest mass energy) photon being
scattered by a direct orbiting electron, 
after which the photon comes
out with a higher energy (inverse Compton scattering). Subsequently,
the target electron may recoil to another direct orbit
of lesser energy,
or the electron may be put on a retrograde orbit of negative energy.
In both cases, the target electron gives up
energy as measured by an observer at infinity. However, to a
particular 
observer in the LNRF this is just a normal 
Compton scattering process in which the
photon loses energy to the electron, 
since the photon arrives at this
local~frame with initial energy higher 
than $\sim \mu_e$ [5].  The infall
 of the
final negative energy electron results in an observer
at infinity measuring a decrease in the 
rotational energy of the KBH.
This inverse Compton scattering process is different from its flat
spacetime counterpart.  In the flat spacetime process, cold
photons are heated by hot electrons to the temperature of the
electrons. In the Penrose process, which occurs in the ergosphere,
generally,  photons
are heated by the rotational energy of the black hole and the
curvature of spacetime [3].

The conditions that the infalling photon encounters
in its free fall through the ergosphere before arriving at the
designated scattering radius are ignored in these calculations.
Otherwise, conditions such as the particle 
density in the ergosphere
and radiative transfer effects would 
have to be incorporated into the
calculations along the null geodesic of the photon.  It is assumed
that these conditions cause little 
qualitative change in the results.
The validity of this assumption can be found in [4], where
it is shown that the variation in the density distribution does 
little
to change the energies of the escaping photons.
The photon arrives at the radius in which the scattering take
place with an initial covariant four-momentum
$(P_{\rm ph})_{\mu}=[(P_{\rm ph})_r,
(P_{\rm ph})_{\Theta},L_{\rm ph},
{-E_{\rm ph}}]$, as measured by an observer at
infinity.  For a photon falling radially inward along
the equatorial plane, $(P_{\rm ph})_{\Theta}=L_{\rm ph}=0$.

The initial energies  $E_{\rm ph}$, used in the 
model calculations presented here, are monochromatic for a given 
distribution
of 2000 incident infalling photons. One scattering 
event per photon is assumed. 
The energies of the incident photons ($0.511$~MeV$-1$~MeV)
are consistent
with the bistable two-temperature accretion disk [8,20,21],
referred to as the thin disk/ion corona 
accretion model (see [5]
for further details concerning the disk 
model and the use of thermal initial
particles).

The spatial distribution of the target electrons 
is assumed to be specified as
circular orbiting rings of completely ionized plasma.
Two radii are considered for the model electron rings:
$r_{\rm ms}$ ($\simeq
1.2M$), the radius of the marginally stable circular orbit
and  $r_{\rm mb}$ ($\simeq 1.089M$),
the radius of the marginally bound
circular orbit, for
a canonical KBH ($a/M=0.998$).  
Note that, the outer ergosphere boundary, the
photon  orbit, and the event horizon of a canonical
KBH are located at 
$r_o = 2M$, $r_{\rm ph} \simeq 1.074M$,
and $r_+\simeq 1.063M$, respectively, in the equatorial plane.  The
region between the radii
$r_{\rm ms}$ and $r_{\rm mb}$ represents the innermost possible
periastrons (radial turning points) for bound
unstable orbiting material particles.
If given a sufficiently large
inward perturbation, a particle in this region 
will eventually fall through the event horizon.
Moreover, $r_{\rm mb}$ is the minimum periastron of all
bound orbits of material particles, before plunging directly
into the black hole.  

After some manipulations of the results given in [5], eq.~(3.47),
 the
final four-momentum components of the PCS photon as measured
by a BLF observer, 
can be expressed by
the following:
\begin{eqnarray}
E_{\rm ph}^\prime&=&\varepsilon_{\rm ph}^{R'}\biggl
	\{\gamma_e(e^\nu+\omega e^\psi
        \beta_\Phi) +\left[\gamma_e e^\nu
	\beta_\Phi+\omega e^\psi
           \left(1+\beta_\Phi^2 {\gamma_e^2\over\gamma_e+1}\right)
          \right] \nonumber \\
       &  &\Bigl(\cos\theta_{\rm ph}^R
       \cos\delta^R+\sin\delta^R
   \sin\theta_{\rm ph}^R\cos\alpha^R\Bigr)\biggr\},
\label{eq:15}
\end{eqnarray}
\begin{equation}
(P_{\rm ph}^\prime)_r=e^{\mu_1}\varepsilon_{\rm ph}^{R'}\sin
                   \theta_{\rm ph}^{R'}
                   \cos\left[\pi-\arccos
		   \left({\cos\delta^R-\cos\theta_
                   {\rm ph}^R\cos\theta_{\rm ph}^{R'}
                    \over\sin\theta_{\rm ph}
                   ^R\sin\theta_{\rm ph}^{R'}}\right)\right],
\label{eq:16}
\end{equation}
\begin{equation}
(P_{\rm ph}^\prime)_\Theta=e^{\mu_2}\varepsilon_{\rm ph}^{R'}
                      \sin\theta_{\rm ph}^{R'}
                     \sin\left[\pi-\arccos\left({\cos\delta^R-
		     \cos\theta_
                      {\rm ph}^{R}\cos\theta_{\rm ph}^{R'}\over
                      \sin\theta_{\rm ph}^R\sin\theta_{\rm ph}^
                       {R'}}\right)
                      \right],
\label{eq:17}
\end{equation}
\begin{equation}
L_{\rm ph}^\prime=e^\psi\varepsilon_{\rm ph}^{R'}
                      \left[\left(1+\beta_\Phi^2{\gamma_e^2
		      \over\gamma_e+1}
                 \right)\Bigl(\cos\theta_{\rm ph}^R\cos\delta^R
		 +\sin\delta^R
                    \sin\theta_{\rm ph}^R\cos\alpha^R\Bigr)
		    +\gamma_e\beta_\Phi
                    \right], 
\label{eq:18}
\end{equation}
where $e^\nu$, $e^\psi$, $e^{\mu_1}$, 
$e^{\mu_2}$, and $\omega$ are given by
Eqs.~(\ref{eq:2}) through (\ref{eq:6}); the other parameters will be 
defined in the following paragraph.  
These four-momentum components
result from scattering off equatorial target electrons (i.e., 
targets confined to the equatorial plane).
See [5] to derive the general expressions 
 for both equatorially and nonequatorially confined
targets.  Note that, the arccosine term in 
Eqs.~(\ref{eq:16}) and~(\ref{eq:17}) defines an angle between
$0$ and $2\pi$.  

The other parameters in the above
four-momentum components [Eqs.~(\ref{eq:15}) through (\ref{eq:18})] 
are defined as follows [5]:  For equatorial orbiting target
electrons, ${\bf \vec\beta}_e=(\beta_r,\beta_\Theta,\beta_\Phi)=
[0,0,(v_e)_\Phi]$, and
$\gamma_e=(1-{\bf \beta}_e^2)^{-1/2}$,
with
\begin{equation}
(v_e)_\Phi=\beta_\Phi={L_e e^{\nu-\psi}\over E_e-\omega L_e},
\label{eq:19}
\end{equation}
where $E_e$ and $L_e$ are the conserved 
energy and azimuthal angular
momentum as measured by an observer at infinity, 
given by Eqs.~(\ref{eq:13})
and~(\ref{eq:14}), respectively, of course with $Q_e=0$ for 
equatorially confined
targets ($Q_e\neq 0$ for nonequatorially confined 
targets); $(v_e)_\Phi$ is
the orbital velocity of the electron relative to the LNRF.  
Note that, in [5], I refer to the lab frame (LF) as the frame of a
general observer in the LNRF; here I simply refer to it as the LNRF.
The final energy of the PCS photon $\varepsilon_{\rm ph}^{R'}$, as 
measured by an observer in the 
electron rest frame (ERF), is given by
the familiar expression for Compton
scattering in flat spacetime (see [5]):
\begin{eqnarray}
\varepsilon_{\rm ph}^{R'}&=&{\varepsilon_{\rm ph}^R
	 \over {1+(\varepsilon_{\rm ph}^R/\mu_e)
			(1-\cos \delta^R)}}\nonumber \\
                  &=&{\gamma_e e^{-\nu}E_{\rm ph}
                   \over 1+\left(\gamma_e e^{-\nu}
                    E_{\rm ph}/\mu_e\right)
                    \left(1-\cos\delta^R\right)},\label{eq:20} 
\end{eqnarray}
where $\varepsilon_{\rm ph}^R$ is the initial energy of the 
photon in the ERF;
$\delta^R$ is the angle between the initial
and final space momentum vectors of the photon in the ERF.
The ERF and the LNRF are related by a 
Lorentz transformation, with relative
frame velocity $(v_e)_\Phi$.
The  initial and final polar
angles, $\theta_{\rm ph}^R$ and $\theta_{\rm ph}^{R'}$, 
of the incoming and scattered 
photon space vectors, respectively, 
in the ERF [in which the coordinate system is centered on the 
electron at rest, with the polar axis 
($+{\bf \hat e_z}$) of this
Lorentz frame pointed in the positive 
${\bf \hat e_\Phi} $ direction
of the BLF; see eqs.~(3.3), (3.5), (3.6), (3.21),
and (3.24) of [5]], are found to be
\begin{eqnarray}
\theta_{\rm ph}^R&=&\arctan\left\{{{\left[(p_{\rm ph}^R)_r^2
                    +(p_{\rm ph}^R)_\Theta^2\right]^{1/2}}
                \over (p_{\rm ph}^R)_\Phi}\right\}\nonumber\\
&=&\pi -\arctan\left({1\over\gamma_e\beta_\Phi}\right)\nonumber\\
&=&\pi -\arctan\left[{2\,m_e^2 M a\, r \sin^2\Theta}\over 
{\omega L_e^2 (r^2+a^2\cos^2\Theta)}\right]^{1/2}
\label{eq:21}
\end{eqnarray}
[using $(p_{\rm ph}^R)_r=-e^{-\nu}E_{\rm ph}$, 
$(p_{\rm ph}^R)_\Theta=0$, 
$(p_{\rm ph}^R)_\Phi=-\gamma_e\beta_\Phi e^{-\nu}E_{\rm ph}$,
and eqs.~(2.3), (2.10b), (2.12),  (3.15) from [5]];
and 
\begin{equation}
\theta_{\rm ph}^{R'}=\arccos(\cos\theta_{\rm ph}^R\cos\delta^R+
\sin\delta^R\sin\theta_{\rm ph}^R\cos\alpha^R)
\label{eq:22}
\end{equation}
[see also the azimuthal angle $\phi_{\rm ph}^{R'}$ [eq.~(3.39)]
of [5], given again here as Eq.~(\ref{eq:44})],
where $(p_{\rm ph}^R)_r$, $(p_{\rm ph}^R)_\Theta$, and 
$(p_{\rm ph}^R)_\Phi$ are the initial space momentum components of the 
photon in the ERF; note, the $\pi$ term has been properly added since  
$(p_{\rm ph}^R)_\Phi<0$.  The angles $\theta_{\rm ph}^{R'}$,
$\phi_{\rm ph}^{R'}$ define the direction of the space momentum vector
of the scattered photon in the ERF (see eq.~(3.24) of [5]). 
The angles $\delta^R$ and $\alpha^R$ are the
 polar and azimuthal angles
of the PCS photon relative to the direction of the initial photon:  
these angles are found by
applying the Monte Carlo method to the Klein-Nishina cross section
in the ERF (as done in [5]). 
Equations~(\ref{eq:21}) 
and~(\ref{eq:22})
will be 
important in the discussions of the GM field, in 
Sections~\ref{sec:2d} and~\ref{sec:3},
because of the frame dragging term $\omega$
contained in $\theta_{\rm ph}^R$
[cf. Eqs.~(\ref{eq:15}) through (\ref{eq:18}), (\ref{eq:19}), 
(\ref{eq:21})], and~(\ref{eq:22})].  

In [5], I present numerous cases of the obtained four-momentum 
spectra of the PCS
photons.  Below in Section~\ref{sec:3}, however, I 
present only spectra that are 
relevant to the discussion at hand---i.e., of how the 
GM field affects the 
distributions of the scattered photons.  Note that, the
conserved energies $E_e$
and angular momenta $L_e$ of the nonequatorially confined target 
electrons,
assumed in these calculations, are  consistent with the electrons 
produced in Eilek's hot accretion disk model [8,20], described
in details in [5].  The $Q_e$ values used in 
Eqs.~(\ref{eq:13}) and~(\ref{eq:14})
for the nonequatorially confined target electrons 
correspond to orbits with polar
latitudinal angles ${<\atop\sim} 30^\circ$, above and below 
the equatorial plane,
consistent with what would be expected in 
thin disk/ion corona (or torus) accretion models.

\subsection{Penrose Gamma Ray-Gamma Ray Pair Production (PPP)}
\label{sec:2c}

This PPP ($\gamma \gamma \longrightarrow e^- e^+$\/) 
process consists of  
collisions inside the ergosphere
between radially infalling photons [5] and circularly orbiting photons
that are bound at the radius of the  photon  orbit [7]: 
\begin{equation}
r_{\rm ph}=2M \lbrace 1+ \cos {[(2/3) \arccos{(-a/M)}]}
\rbrace,
\label{eq:23}
\end{equation}
 where $r_{\rm ph}$ represents an
unstable circular orbit.  On such spherical-like orbits, a photon
circles indefinitely; however, a small perturbation can cause the
photon to either escape to infinity or fall into the black hole. 
This orbit is
the innermost boundary of circular orbits for all particles.
 The initial energies assumed in these
calculations as measured by an observer at infinity, for the
infalling photon and the orbiting photon, 
$E_{\gamma 1}$ and $E_{\gamma 2}$,
respectively, should be in the following ranges to be 
consistent with energies produced by thin disk/ion corona
models, and with energies satisfying the threshold 
energy conditions for these reactions to occur [5]:
$$
3.5 {\rm~keV}\,{<\atop\sim}\, 
E_{\gamma 1}\,{<\atop\sim}\, 50~{\rm keV}~~~
{\rm and}~~~3.4~{\rm MeV}\,{<\atop\sim}\, E_{\gamma 2}\,{<\atop\sim}\, 
4~{\rm GeV},
$$
where $E_{\gamma 2}$ is given by Eq.~(\ref{eq:13}) 
for the appropriate chosen 
value of $Q_{\gamma 2}$.  
Note that, the target photons, $E_{\gamma 2}$, 
have orbits not confined to the 
equatorial plane.
These nonequatorially confined orbits allow for the 
photon's energy at the photon
orbit to be used in a feasible way, whereas, for 
equatorially confined
orbits, Eqs.~(\ref{eq:13}) and~(\ref{eq:14}), 
for $E$ and $L$, respectively,
equal zero [cf.~Fig.~1(a)].  Or expressed
in conventional terms, $E$ and $L$, when $Q$=0, define the orbit of 
infinite energy and infinite azimuthal angular momentum per unit rest
mass energy, 
respectively [7,5].

The high energy range for $E_{\gamma 2}$ is
chosen based on physical processes found to 
occur in the two-temperature bistable thin disk/ion
corona accretion flow.  More recently such type 
ion corona accretion
flows have been called  advection
dominated accretion flows (ADAFs).  ADAFs were predicted in Lightman's
speculation of what might take place as a result of 
thermal and density instabilities [22]. These instabilities
can cause
the thin disk to form high and low density rings in the inner
region and an increase
in temperature from $\sim 10^{5-7}$~K to $\sim 10^{9-12}$~K
[8,20,21,22].
Astrophysicists over the years, since
Lightman's time-dependent  disk model
calculation of the early 1970s [22],
and subsequent models [8,20], 
have continued to find properties of ADAFs [23].
The most
important is the proton-proton collisions in which neutral pions
$\pi^\circ$ are emitted [8,20,23].
The neutral pion $\pi^\circ$ subsequently decays into two energetic
$\gamma$-rays with energies up to $\sim 100$~MeV  if the protons
are assumed to have a blackbody  distribution, or higher for
an assumed power-law distribution [23]. 
Such photons could populate the photon orbit as follows.  
These infalling photons
would be blueshifted in energy-momentum
at the photon orbit by a
factor  $\sim e^{-\nu}\simeq 52$
(due to the rotating black hole producing the Lense-Thirring effect),
where $e^{-\nu}$ is the  so-called blueshift factor given by 
Eq.~(\ref{eq:2}) [cf.~eq.~(2.8d) of [5]].
This means that high energies
up to $\sim 5$~GeV (or higher) 
can possibly exist in the ergosphere at or near the photon orbit.
However, it
requires a process, such as the Penrose mechanism, to inject  these
highly blueshifted {\it trapped} orbiting 
photons into escaping orbits [7]---in 
this PPP ($\gamma \gamma \longrightarrow e^- e^+$), they escape
in the form of energetic $e^-e^+$ pairs.  
Now,
it is well known  that for rotating black holes with $a/M=0.998$,
the accretion disk extends inward to $\sim r_{\rm ms}$
($\simeq 1.2M$), inside of the
radius of the
ergosphere ($r_o=2M$). 
So, the  $\gamma$~rays produced in Eilek's
hot accretion disk model [8,20], from the decay of neutral
pions $\pi^\circ\longrightarrow\gamma\gamma$,
described in details in [5],
with energies peaked around $\sim 75 $~MeV, could very well
be seed photons for
these $\gamma \gamma \longrightarrow e^- e^+$\/\space 
reactions, photons
that have become blueshifted and
bound at the photon orbit
[24]. In addition, the PCS
processes
of [5] show that some of the scattered photons can also be
seed particles for these reactions.
Further, these photon collisions
can occur at other radii besides the
photon orbit, at, say another  turning
point, thereby increasing the probability of a collision.
Again, it is important to point out that once the photons have been
blueshifted to energies given by $E_{\gamma 2}$, and become
bound at the photon orbit,
they have no other way
of escaping, except by some perturbing physical
interaction, such as these
PPP ($\gamma \gamma \longrightarrow e^- e^+$\/) processes.  
Note that, in [5], these PPP ($\gamma \gamma \longrightarrow 
e^- e^+$\/) processes are
defined as  $quasi$-Penrose processes since they are somewhat
different from classical Penrose processes---which utilize the
retrograde orbits of particles in the ergosphere.

Before proceeding, I want to discuss the assumptions made above 
and their 
validities.  First, it is assumed that the blueshift in the energy
of an infalling orbiting photon
as measured by a LNRF observer is approximately equal to that 
measured by a BLF observer.  This is a valid assumption 
because, for a photon
in a  circular bound orbit about the KBH, there is little difference
between the energy measured locally in the LNRF [eq.~(2.8d) 
of [5]] and that 
measured
by an observer at infinity (i.e., in the BLF) 
[Eq.~(\ref{eq:13})]; for
example,  at the photon orbit $r_{\rm ph}$, the ratio of the
energy measured by an observer in the LNRF, 
$\varepsilon_{\rm ph}^*$,
to that measured by an observer in BLF, $E_{\rm ph}^*$, is 
$\simeq 1.055$.  Second, it is assumed that the photons
populating the photon orbit have been created by some
prior processes (disk instability and/or Penrose), which yielded
them with appropriate energies and momenta.  We know from 
[5]  that the condition
\begin{displaymath} 
\eta_{\rm ph}^{\prime} \geq \eta_{\rm ph}^*\equiv
{Q_{\rm ph}^{\prime} \over E_{\rm ph}^{{\prime}
^2}} \geq{Q_{\rm ph}^*
\over E_{\rm ph}^{*^2}}
\end{displaymath}
guarantees  the existence of a turning point 
outside the event horizon for which the radial motion goes 
to zero and changes direction for photons  emitted not confined
to the equatorial plane
[i.e., $(P_{\rm ph}^\prime)_\Theta\neq 0$].  
The asterisks
indicate the nonequatorially confined circular 
photon orbit, where  $r=r_{\rm ph}$ [see Eq.~~(\ref{eq:23})], 
which is a potential 
turning point, and the primes indicate a photon resulting 
from some 
prior process, say PCS.  Once  the condition above 
has been met, indicating the existence of a turning point
outside the horizon,
to find out whether or not the photon orbit is a true  
turning point, for a particular photon,  it seems plausible
to  set $Q_{\rm ph}^{\prime}=Q_{\rm ph}^*$, in the 
 the condition above, and let $E_{\rm ph}^\prime
\longrightarrow e^{-\nu}E_{\rm ph}^\prime\equiv E_{\rm ph}^{\prime
\prime} $,  as the photon spirals inward toward
the photon orbit,
  obtaining the requirement  
that  $E_{\rm ph}^{\prime\prime}
\leq E_{\rm ph}^{*}$ for the existence of a turning point 
at $r=r_{\rm ph}$ [where
$e^{-\nu}$ is the gravitational blueshift factor, again, 
given by Eq.~(\ref{eq:2}),
due to the local BLF orbital energy gain; cf. also Eq.~(\ref{eq:1})]. 
  After the photon has reached its turning 
point, it seems reasonable that it escapes outward with ``redshifted'' energy 
given by 
$E_{\rm ph}^\prime=
e^\nu E_{\rm ph}^{\prime\prime}$ as measured by an  
observer at infinity, i.e., in the BLF.  Although at this point we do 
not know the exact detailed path of the photon to and from its turning point, 
 but, we can theorize that it looses its 
acquired inspiraling local orbital energy on its way back outward, before
escaping to infinity, as indicated in the above relationship between 
$E_{\rm ph}^\prime$ and $E_{\rm ph}^{\prime\prime}$.
Further, notice that 
the condition above is 
independent of $L_{\rm ph}^\prime$.  The other condition that 
independently
guarantees the existence of a turning point in $P_r$,
however,  for photons  confined 
to escape along the equatorial plane, as well as for those escaping
not confined to the  equatorial plane, 
involves $L_{\rm ph}^\prime$;
see [5] for  details and a
complete description of the escape conditions.
In [5], it is found that many of the PCS
photons have appropriate $E_{\rm ph}^{\prime}$
and  $Q_{\rm ph}^{\prime} $ values
to populate the turning point trajectories 
at or near the photon orbit [cf.~Figs. 3 and~4 to~Fig. 1(a)].
Because of this, it is not unreasonable to expect the photons
produced by the $\pi^o \longrightarrow\gamma\gamma$ decays
(discussed above) 
to acquire appropriate
$E_{\rm ph}^{\prime}$
and  $Q_{\rm ph}^{\prime} $ values, particularly, if the process,
starting from the proton-proton scattering,  is properly
considered as a Penrose process inside the ergosphere.
Note,  the radii of other turning points can be found 
from Eqs.~(\ref{eq:13}) and~(\ref{eq:14}), by  satisfying the 
condition 
 $ E_{\rm ph}^{\prime\prime}\equiv e^{-\nu}E_{\rm ph}^{\prime}
\leq E_{\rm ph}^{*}$, 
for the specified
``iso-Q'' value orbit ($\,\equiv\,$circular orbit of equal $Q$),
where now the radius of the turning point can exist 
anywhere near the event horizon,  say between
$r_{\rm mb}>r>r_{\rm ph}$ for the inward PCS photons.   
This means that such photons  
satisfying a condition for the existence of a turning point
can have energies at these turning points corresponding to
 blueshift
factors in the range of 
$32<e^{-\nu}<52$.  The largest factors,  
however, will be for photons nearest the photon orbit.  
In these calculations,
the investigation centers around those photons with turning
points at the photon orbit where the highest energy can be 
extracted.   Yet, consideration of all of the possible 
turning points will increase the probability 
of a PPP ($\gamma \gamma \longrightarrow e^- e^+$\/) 
process occurring, 
as mentioned earlier, giving rise to a larger
volume of $e^-e^+$ pairs.   Moreover,   
no null geodesic 
has more
than one turning point in its radial motion
 outside the event horizon.  In general, a bound orbit at constant 
radius is considered as an orbit with a ``perpetual'' turning point.
The bound unstable orbit of the photon at constant radius
(the {\it photon orbit}\/) may be considered as an orbit on the verge 
of having a perpetual turning point.   
In addition, before proceeding, one way to described
particles trajectories in the ergosphere is the following:
As a particle travels with $P_r>0$ or $P_r<0$ to, say $P_r=0$, 
all local inertial 
frames as measured by an observer at infinity, i.e., in the BLF,
are dragged into rotation by the KBH, irrespective of the 
four-momentum of the particle as it moves relative to local
or neighboring inertial frames. (Note, particles could also be 
undergoing local 
relativistic scatterings as in the Penrose processes considered
here.)
  The BLF observer measures the effects, given by 
$P_\mu=(P_r,P_\Theta,L,-E)$,  of
this frame dragging  on the particle as it moves in such a 
strong, locally constant,  gravitational environment 
(more on this in 
Section~\ref{sec:3a}).

Now proceeding, manipulations of the results given in [5] yield,
in the following form,
the final four-momentum  components
of the PPP positron ($e^+$) and  negatron ($e^-$)
as measured by an observer in the BLF:
\begin{eqnarray}
E_+&=&e^\nu\gamma_{\rm c.m.}\biggl\{\varepsilon
     _+^c+\left(\varepsilon_+^{c^2}-\mu_e^2
     \right)^{1/2}\Bigl[(\beta_{\rm c.m.})_r\sin\theta_{e^+}^c
      \cos\phi_{e^+}^c\nonumber\\
      & &+ (\beta_{\rm c.m.})_\Theta\sin\theta_{e^+}^c
       \sin\phi_{e^+}^c
        +(\beta_{\rm c.m.})_\Phi\cos\theta_{e^+}^c\Bigr]\biggr\}
       +\omega e^\psi\Biggl[\left(\varepsilon_+^{c^2}
       -\mu_e^2\right)^{1/2}\nonumber\\
       & &\Biggl\{{\gamma_{\rm c.m.}^2\over
       \gamma_{\rm c.m.}+1}
        (\beta_{\rm c.m.})_\Phi
       \Bigl[(\beta_{\rm c.m.})_\Theta
         \sin\theta_{e^+}^c\sin\phi_{e^+}^c\nonumber\\
      & & +(\beta_{\rm c.m.})_r
      \sin\theta_{e^+}^c\cos\phi_{e^+}^c\Bigr]
       +\Biggl[1+(\beta_{\rm c.m.})_
        \Phi^2{\gamma_{\rm c.m.}^2\over\gamma_
       {\rm c.m.}+1}\Biggr]\nonumber\\
       & &\cos\theta_{e^+}^c
       \Biggr\}+\gamma_{\rm c.m.}(\beta_{\rm c.m.})_
         \Phi\varepsilon_+^c\Biggr],\label{eq:24}
\end{eqnarray}
\begin{eqnarray}
(P_+)_r&=&e^{\mu_1}\Biggl[\left(\varepsilon_+^{c^2}
-\mu_e^2\right)^{1/2}
\Biggl\{{\gamma_{\rm c.m.}^2\over\gamma
_{\rm c.m.}+1}(\beta_{\rm c.m.})_r
\Bigl[(\beta_{\rm c.m.})_\Phi\cos\theta_{e^+}^c\nonumber\\
& &+(\beta_{\rm c.m.})_\Theta\sin\theta_{e^+}^c\sin\phi_{e^+}^c\Bigr]
+\Biggl[1+(\beta_{\rm c.m.})_r^2{\gamma
_{\rm c.m.}^2\over\gamma_
{\rm c.m.}+1}\Biggr]\nonumber\\
& &\sin\theta_{e^+}^c\cos\phi_{e^+}^c\Biggr\}
+\gamma_{\rm c.m.}(\beta_{\rm c.m.})_r\varepsilon_+^c
\Biggr],\label{eq:25}
\end{eqnarray}
\begin{eqnarray}
(P_+)_\Theta&=&e^{\mu_2}\Biggl[\left(\varepsilon_+^{c^2}
-\mu_e^2\right)^{1/2}
\Biggl\{{\gamma_{\rm c.m.}^2\over\gamma
_{\rm c.m.}+1}(\beta_{\rm c.m.})_\Theta
\Bigl[(\beta_{\rm c.m.})_\Phi\cos\theta_{e^+}^c\nonumber\\
& &+(\beta_{\rm c.m.})_r
\sin\theta_{e^+}^c\cos\phi_{e^+}^c\Bigr]
+\Biggl[1+(\beta_{\rm c.m.})_\Theta^2{\gamma
_{\rm c.m.}^2\over\gamma_
{\rm c.m.}+1}\Biggr]\nonumber\\
& &\sin\theta_{e^+}^c\sin\phi_{e^+}^c\Biggr\}
+\gamma_{\rm c.m.}(\beta_{\rm c.m.})_\Theta\varepsilon_+^c\Biggr],
\label{eq:26}
\end{eqnarray}
\begin{eqnarray}
(P_+)_\Phi=L_+&=&e^\psi\Biggl[\left(\varepsilon_+^{c^2}-
\mu_e^2\right)^{1/2}
       \Biggl\{{\gamma_{\rm c.m.}^2\over\gamma_{\rm c.m.}+1}
        (\beta_{\rm c.m.})_\Phi\nonumber\\
       & &\Bigl[(\beta_{\rm c.m.})_\Theta
       \sin\theta_{e^+}^c\sin\phi_{e^+}^c
       +(\beta_{\rm c.m.})_r\sin\theta_{e^+}^c\cos\phi_{e^+}^c\Bigr]
       \nonumber\\
       & &+\Biggl[1+(\beta_{\rm c.m.})_\Phi^2{\gamma
       _{\rm c.m.}^2\over\gamma_
       {\rm c.m.}+1}\Biggr]\cos\theta_{e^+}^c
       \Biggr\}\nonumber\\
       & &+\gamma_{\rm c.m.}(\beta_{\rm c.m.})_\Phi\varepsilon_+^c
        \Biggr],
\label{eq:27}
\end{eqnarray}
where, again, $e^\nu$, $e^\psi$,
$e^{\mu_1}$, $e^{\mu_2}$, and $\omega$
are given by
Eqs.~(\ref{eq:2}) through (\ref{eq:6}), however,
evaluated at the photon orbit
$r_{\rm ph}$ [Eq.~(\ref{eq:23})].  Note, sometimes I refer to the
$e^-e^+$ pairs as PPP electrons.

The other expressions in the above
four-momentum components
[Eqs.~(\ref{eq:24}) through (\ref{eq:27})]
are defined as follows
(just as in the PCS,
these definitions are derived from [5]; the results are
presented here).
The subscript c.m. defines the center-of-momentum frame,
and the superscript $c$ indicates parameters
measured relative to this frame.
The c.m.~frame and the LNRF are related by a Lorentz transformation,
with relative frame  velocity defined by
${\vec\beta}_{\rm c.m.}={\bf v}_{\rm c.m.}=[(\beta_{\rm c.m.})_r,
(\beta_{\rm c.m.})_\Theta,(\beta_{\rm c.m.})_\Phi]$, and
with the Lorentz factor of the c.m. frame given by
$\gamma_{\rm c.m.}=
(1-\beta_{\rm c.m.}^2)^{-1/2}$.  In general,
\begin{equation}
{\vec\beta}_{\rm c.m.}={{{\bf p}_{\gamma 1}+{\bf p}_{\gamma 2}}\over
            {\varepsilon_{\gamma 1}+\varepsilon_{\gamma 2}}},
\label{eq:28}
\end{equation}
where $\varepsilon_{\gamma 1}$
and ${\bf p}_{\gamma 1}$ are the energy and space
momentum vector, respectively,  of the radially infalling
photon as measured by a LNRF
observer;  $\varepsilon_{\gamma 2}$ and
${\bf p}_{\gamma 2}$ are these
measured quantities for the orbiting photon.  Using
eqs.~(3.90)-(3.92),
with $\theta_{\gamma 1}=\pi/2$ and $\phi_{\gamma 1}=\pi$,
and (2.8a)-(2.8c) of [5] in Eq.~(\ref{eq:28}) above, and
after some manipulation,
we find that
\begin{eqnarray}
{\vec\beta}_{\rm c.m.}&=&[(\beta_{\rm c.m.})_r,
(\beta_{\rm c.m.})_\Theta,
(\beta_{\rm c.m.})_\Phi]\nonumber\\
&=&(E_{\gamma 1}+E_{\gamma 2}-\omega L_{\gamma 2})^{-1}
(-E_{\gamma 1},
e^{\nu-\mu_2}\sqrt{Q_{\gamma 2}},e^{\nu-\psi}L_{\gamma 2}),
\label{eq:29}
\end{eqnarray}
where $E_{\gamma 1}$ and $E_{\gamma 2}$ are the conserved
initial energies
of the infalling incident and target
orbiting photons, respectively,
measured in the BLF,
and $L_{\gamma 2}$ is the
corresponding conserved azimuthal angular
momentum of $E_{\gamma 2}$.
The c.m. frame energy is given by [5]
\begin{eqnarray}
\varepsilon_+^c&=&\sqrt{{1\over 2}\varepsilon_{\gamma 1}
\varepsilon_{\gamma 2}
(1-\cos\theta_{\gamma \gamma})}\nonumber\\
&=&e^{-\nu}\sqrt{{1\over 2}E_{\gamma 1}
         (E_{\gamma 2}-\omega L_{\gamma 2})},
\label{eq:30}
\end{eqnarray}
where $\theta_{\gamma \gamma}~(=\pi/2)$ is the polar
angle between ${\bf p}_{\gamma 1}$ and
${\bf p}_{\gamma 2}$ in the LNRF.
The polar and azimuthal angles ($\theta_{e^+}^c,
\phi_{e^+}^c$) of the $e^+$
in the c.m. frame, with the polar axis of
the coordinate system pointed in
the positive ${\bf \hat e_\Phi}$ direction,
are given below in terms of the
scattering angles ($\theta_+^c, \phi_+^c$),
which are found from application
of the Monte Carlo method to the cross
section for these pair production
scattering events [5]:
\begin{eqnarray}
\theta_{e^+}^c&=&-\arcsin{\left({\cos\theta_{+}^c\over
            \cos\phi_{e^+}^c}\right)};\label{eq:31}\\
\phi_{e^+}^c&=&\arctan{\left[{{-D_2+\left({D_2^2-4D_1 D_3}\right)
^{1/2}}\over{2D_1}}\right]}, \label{eq:32}
\end{eqnarray}
where
\begin{eqnarray*}
D_1&\equiv&\cos^2\theta_{+}^c\sin^2\theta_{+}^c; \\
D_2&\equiv& 2\cos\theta_{+}^c\sin\theta_{+}^c\sin\phi_{+}
^c(1-\cos^2\theta_{+}^c);\\
D_3&\equiv& 1+\cos^2\theta_{+}^c(\cos^2\theta_{+}^c-2)
     +\sin^2\theta_{+}^c\cos^2\phi_{+}^c(\cos^2\theta_{+}^c -1).
\end{eqnarray*}
The corresponding angles for the $e^-$ are given by
\begin{eqnarray}
\theta_{e^-}^c&=&\pi-\theta_{e^+}^c,\label{eq:33}\\
\phi_{e^-}^c&=&\phi_{e^+}^c  + \pi.\label{eq:34}
\end{eqnarray}
Note that, to find the corresponding four-momentum
components of Eqs.~(\ref{eq:24}) through (\ref{eq:27}) for the $e^-$,
change the subscripts to ``$-$'',
using the fact that $\varepsilon_+^c=\varepsilon_-^c$, and let
$\theta_{e^+}^c\longrightarrow\theta_{e^-}^c$,
$\phi_{e^+}^c\longrightarrow\phi_{e^-}^c$,
with the new angles defined by Eqs.~(\ref{eq:33})
and~(\ref{eq:34})
above.
Equations~(\ref{eq:29})
and~(\ref{eq:30})
will be
important in the discussion of the GM field in
Sections~\ref{sec:2d} and~\ref{sec:3},
i.e., because of the frame dragging term $\omega$
contained in these equations
[cf. Eqs.~(\ref{eq:24}) through (\ref{eq:27}), (\ref{eq:29})
and~(\ref{eq:30})].

Similarly, as for the PCS photons, numerous cases of the
obtained four-momentum spectra of the PPP
($\gamma \gamma \longrightarrow e^- e^+$)
electrons are presented in [5].
Below, in Section~\ref{sec:3}, I present only spectra that are
relevant to the discussion at hand of how
the GM field affects the Penrose produced $e^-e^+$ pairs.

\subsection{The Gravitomagnetic  (GM) Field}
\label{sec:2d}

In order to understand and to appreciate the origin of the
asymmetry in the distributions
of the scattered particles in the polar ($\Theta$) direction,
mentioned in Section~\ref{sec:1},
one must look in details at the gravitomagnetic (GM)
force field [10]---the force
responsible for  the asymmetry.
The gravitational force of a rotating
body of mass consists of two parts,
which are analogous to an electromagnetic field.
The first part is the familiar gravitational
force that a body of rotating or nonrotating mass $M$ produces
on a test particle of mass $m$:
\begin{equation}
\Bigg\lgroup{d{\bf p}\over d\tau}\Bigg\rgroup_{\rm grav}
={m\over (1-{\bf v}^2)^{1/2}}\,{\bf g},
\label{eq:35}
\end{equation}
where the gravitational acceleration $\bf g$ is
produced by the gradient of
the ``redshift factor'' $e^\nu$ of Eq.~(\ref{eq:2});
note, the inverse $e^{-\nu}$ is referred to as the blueshift
factor [5,9]:
\begin{eqnarray}
{\bf g}&=&-\nabla \ln e^\nu \nonumber\\
       &=&-{\Sigma M\bigl(r^4-a^4\bigr)
       +2Mr^2a^2\Delta\sin^2\Theta\over
          A\sqrt{\Delta\Sigma^3} }\,{\bf \hat e_r} \nonumber\\
        &~~~&+{2Mra^2\bigl(r^2+a^2\bigr)\over A\sqrt{\Sigma^3}}
           \cos\Theta\sin\Theta\,{\bf \hat e_\Theta}.
\label{eq:36}
\end{eqnarray}
The above force is analogous to the electric
force field surrounding an
electric charge source (i.e., like the Coulomb
field between point charges),
and for this reason is sometimes referred to as the
``gravitoelectric'' force.

The second part, however, is
less familiar. It is the additional
gravitational force that a rotating mass
produces on a test particle.  This force, called the GM force,
is produced by the gradient of
${\vec \beta_{_{\rm GM}}}= -\omega{\bf \hat e_\Phi}$,
where $\omega$ ($=-g_{\Phi t}/g_{\Phi\Phi}$ [7]) is the frame
dragging velocity appearing
in Eq.~(\ref{eq:1}), expressed by
Eq.~(\ref{eq:6}).  The GM force exerted on a test particle
of space momentum ${\bf p}$
is given by [10]
\begin{equation}
\Bigg\lgroup{d{\bf p}\over d\tau}\Bigg\rgroup_{\rm GM}=
{\bf
\tensor{\bf H} \cdot p}, \;\;{\rm i.e.,}\;
\Biggl(\Bigg\lgroup{d{\bf p}\over d\tau}
\Bigg\rgroup_{\rm GM}\Biggr)_{i}
=H_{ij}p^{j},
\label{eq:37}
\end{equation}
with
\begin{displaymath}
\tensor{\bf H}\equiv e^{-\nu}\nabla\vec\beta_{_{\rm GM}},
                    \;\;{\rm i.e.,}\;
H_{ij}=e^{-\nu}(\beta_{_{\rm GM}})_{_{j\vert i}}
\end{displaymath}\
(the vertical line indicates the covariant derivative in
3-dimensional absolute space).
The field $\tensor{\bf H}$
 is called the
GM tensor field, and
$\vec \beta_{_{\rm GM}}$ is sometimes called the GM potential.
The components of $\tensor{\bf H}$
 in the LNRF are [10]
\begin{eqnarray*}
H_{\Theta\Phi}&=&-{2aMr\cos\Theta\over\Sigma^2\,\sqrt{A^3}}
              \Bigl[\bigl(r^2+a^2\bigr)A+a^2\Sigma\Delta
              \sin^2\Theta\Bigr]e^{-\nu}\equiv H^r,\\[.25in]
H_{\Phi\Theta}&=&{2aMr\cos\Theta\over\Sigma^2\,\sqrt{A^3}}
              \Bigl[\bigl(r^2+a^2)A-a^2\Sigma\Delta\sin^2
                               \Theta \Bigr]e^{-\nu}
                               \equiv-\tilde H^r,
\end{eqnarray*}
\begin{eqnarray}
H_{r\Phi}&=&-{2aM\sqrt{\Delta}\sin\Theta\over\Sigma^2\, \sqrt{A^3}}
                \Biggl\{a^2\cos^2\Theta A-\Sigma r\Bigl[2r
                \bigl(r^2+a^2\bigr)\nonumber\\
                & & -a^2\bigl(r-M\bigr)\sin^2\Theta\Bigr]
          \Biggr\}e^{-\nu}\equiv-\tilde H^\Theta,\nonumber\\[.25in]
H_{\Phi r}&=&-{2aM\sqrt{\Delta}\sin\Theta\over\Sigma^2\, \sqrt{A^3}}
                                \Biggl\{r^2A-\Sigma r\Bigl[2r\bigl(
                             r^2+a^2\bigr)\nonumber\\
                             & &-a^2\bigl(r-M\bigr)\sin^2
                                \Theta\Bigr]\Biggr\}e^{-\nu}
                                 \equiv H^\Theta.
\label{eq:38}
\end{eqnarray}
(The above definitions will become clearer in later sections.)
Note that, like the magnetic Lorentz force ${\bf f}_{_B}=q({\bf v}
\times{\bf B})$, on a particle of charge $q$, moving with velocity
$\bf v$, in a magnetic field $\bf B$,
the GM force of Eq.~(\ref{eq:37}) vanishes when the particle is at rest.

The justification for the word
``gravitomagnetic''  becomes clear when
we look at the GM force
at a large distance
from the event horizon
($r\gg r_+$), exerted on a particle of mass m
traveling with low velocity ($v\ll c$).
The GM force as measured by an observer at infinity [of whom I
will refer to, after the introduction of Eq.~(\ref{eq:47}), as the
asymptotic rest BLF observer],
under these conditions, is given by
${\bf F}_{_{\rm GM}}\simeq m({\bf v}\times{\bf H})$,
where $\bf H$ is
called the GM vector field---which contains the same information
as the antisymmetric part of the GM tensor field [10]:
\begin{displaymath}
H^j\equiv\epsilon^{jkl}H_{kl}
=\epsilon^{jkl}e^{-\nu}\bigtriangledown_k
(\beta_{_{\rm GM}})_{_{l}}\,;\;\;{\rm i.e.,} 
~{\bf H}\equiv e^{-\nu}
\nabla\times \vec\beta_{_{\rm GM}}, 
\end{displaymath}
or
\begin{eqnarray}
{\bf H}&=&H^\Theta{\bf \hat e_\Theta} + H^r {\bf \hat e_r} 
            \nonumber\\
        &=&-{2aM\over\sqrt{\Sigma^5}}
	 \Biggl[\bigl(r^2-a^2\cos^2\Theta\bigr)
        \sin\Theta{\bf \hat e_\Theta} + {2r\bigl(r^2+
	a^2\bigr)\over\sqrt{\Delta}}
        \cos\Theta{\bf \hat e_r} 
\Biggr].
\label{eq:39}
\end{eqnarray}
Thus, one sees that the GM force on a test particle of
mass $m$ is similar to the Lorentz magnetic force ${\bf f}_{_B}$ 
on a test particle
of charge $q$, as given above, with the vectors $\bf H$ and $\bf B$
being analogous.  In addition, the lines of force of $\bf H$
have characteristics similar to that 
of a dipole magnetic field.  Therefore, because of
resemblances such as these (more examples given in [10]),
 of the force field produced by the mass and angular momentum 
of the rotating black hole,  to that of a magnetic force field, 
thence came 
the term gravitomagnetic.
  
To assure a further understanding of the GM force field,
one can think of it 
as the following. Just as the gravitational force causes
local inertial frames to fall with acceleration $\bf g$, the GM
field $\bf H$ causes local inertial 
frames to rotate with angular velocity 
$\overrightarrow \Omega_{_{\rm GM}}=-{\bf H}/2$.  In other words, 
the GM field
$\bf H$ can be thought of as a force field that 
drags local inertial
frames into rotation and, as a result, 
produces a ``Coriolis force''
$m({\bf v}\times\bf H)$ at $r\gg r_+$.  The angular rotation velocity
$\Omega_{_{\rm GM}}$ 
is universally induced in all bodies and neighboring
inertial frames, at a given radius, i.e., independent of the 
mass $m$ of the body, just as in the case of $\bf g$.   

Moreover, $\overrightarrow \Omega
_{_{\rm GM}}$ is equal to the precession 
rate of the spin $\bf s$ of a gyroscope in the GM field of a rotating
body [10].  The angular momentum 
of the gyroscope is changed by the torque
${d{\bf s}/dt}=(1/2){\bf s} \times{\bf H}$, analogous to 
the torque $\vec \mu\times{\bf
 B}$ exerted by a magnetic field $\bf B$ on a magnetic dipole moment
$\vec\mu$ of a ``test'' magnet.  Also, the precession rate 
$\overrightarrow \Omega_{_{\rm GM}}$, like the acceleration  
of gravity
$\bf g$, is independent of the composition and 
structure of the gyroscope.
It is a universal precession induced in all bodies, 
even those with vanishingly
small angular velocity.

The effect of inertial frame dragging, specifically that concerning
the GM precession of gyroscopes, was discovered 
in the equations of general 
relativity by Thirring and Lense in 1918 [1] 
and, thus, is often called
the Lense-Thirring effect. Since there has 
not yet been a detection of the 
GM field of any rotating body [10]---and to date (1997, original
copyright date of this manuscript) this still may 
be the case, it would be another significant 
test of general relativity, if we find, as these calculations 
suggest, that the one-sidedness in the jets 
of AGNs is due largely to
the GM field.

I now use Eqs.~(\ref{eq:37}) and~(\ref{eq:38}) to evaluate the 
force experienced by a test particle
under the influence of the GM field, as measured by a LNRF observer.
In the LNRF, since the covariant components 
of a four-momentum vector
$\lbrace p_\mu=[p_r,p_\Theta,p_\Phi,p_t~
(=- \varepsilon)]\rbrace$  are related to the 
contravariant components by $p^\mu
=\eta^{\mu \nu}p_\nu$, where $\eta^{\mu \nu} \equiv\,$  
Minkowski metric
components, then the space components of the GM force, exerted on
a test particle, are found to be
\begin{eqnarray}
\Bigl(f_{_{\rm GM}}\Bigr)_r&=&-{2aM\sqrt{\Delta}
		       \sin\Theta\over\Sigma^2\, 
                        \sqrt{A^3}} 
                           \Biggl\{a^2\cos^2\Theta A-\Sigma r\Bigl[2r
                           \bigl(r^2+a^2\bigr) \nonumber\\
                           & &-a^2\bigl(r-M\bigr)\sin^2\Theta\Bigr] 
                     \Biggr\}p_\Phi\equiv -\tilde H^\Theta p_\Phi,
\label{eq:40}
\end{eqnarray}
\begin{equation}
\Bigl(f_{_{\rm GM}}\Bigr)_\Theta=-{2aMr\cos\Theta\over\Sigma^2\,\sqrt{A^3}}
                              \Bigl[\bigl(r^2+a^2\bigr)A+a^2
			      \Sigma\Delta
                         \sin^2\Theta\Bigr]p_\Phi\equiv H^r p_\Phi,
\label{eq:41}
\end{equation}
and
\begin{eqnarray}
\Bigl(f_{_{\rm GM}}\Bigr)_\Phi&=&-{2aM\sqrt{\Delta}
			       \sin\Theta\over\Sigma^2\, 
                   \sqrt{A^3}}
                                \Biggl\{r^2A-\Sigma r\Bigl[2r\bigl(
                            r^2+a^2\bigr)\nonumber\\
			    & &-a^2\bigl(r-M\bigr)\sin^2
                                \Theta\Bigr]\Biggr\}p_r
                             +{2aMr\cos\Theta
			     \over\Sigma^2\,\sqrt{A^3}}
			     \nonumber\\
                            &  &\Bigl[\bigl(r^2+a^2)A-a^2
			    \Sigma\Delta\sin^2
                              \Theta \Bigr]p_\Theta
                          \equiv H^\Theta p_r-\tilde H^r p_\Theta.  
\label{eq:42}
\end{eqnarray}
(Again, the definitions will become clearer in later sections.)
Even though these GM force components 
are what is measured locally, i.e.,
in the LNRF, they are sufficient for our purposes to describe what 
we observe at infinity, i.e., in the BLF, 
since the space momentum components in the
LNRF ($p_r,p_\Theta,p_\Phi$) and the BLF ($P_r,P_\Theta,P_\Phi$) are 
related in a linear fashion (see the transformations 
between these frames
given by eqs.~(2.7) and~(2.8) of [5]). 
Notice that the radial component $(f_{_{\rm GM}})_r$ and the polar 
component $(f_{_{\rm GM}})_\Theta$ [Eqs.~(\ref{eq:40}) and~(\ref{eq:41})],
exerted by the GM field, are proportional to the azimuthal
momentum of the particle.  
This means that the absolute magnitudes of these
force components will increase as the energy, $E^\prime$,
 of the scattered
particle increases, since $P_\Phi^\prime$ 
($= L^\prime$) and $E^\prime$
increase linearly [5].  The azimuthal  component
$(f_{_{\rm GM}})_\Phi$ of Eq.~(\ref{eq:42}) will act according to
the radial and polar momenta of the incident and scattered particles,
exerting a force in the azimuthal ($\Phi$) direction. 
This GM force component could be important in the polar 
collimation effects produced by the rotating
black hole [25,26]; however, it is beyond the scope of this 
manuscript to discuss it in that context here.

Displayed in Figs.~2(a) and 2(b) are the 
ratios $(f_{_{\rm GM}})_r/p_\Phi$
and $(f_{_{\rm GM}})_\Theta/p_\Phi$, respectively, as functions of the 
polar angle $\Theta$, which will be compared 
to the polar angle of escape
for a Penrose scattered particle. 
The angle $\Theta$ varies from $0$ to $\pi$, 
with the equatorial plane located 
at $\pi/2$.   
For a given positive $p_\Phi$, as opposed to negative $p_\Phi$ for 
retrograde motion, notice  
that  $(f_{_{\rm GM}})_r$ of Fig.~2(a)
reaches a maximum in the equatorial plane, 
and increases with increasing
$r$ away from the event horizon, within the range of $r$ considered.  
In Fig.~2(b), we notice that 
$(f_{_{\rm GM}})_\Theta$ is antisymmetric, above 
and below
the equatorial plane,
and  reaches  negative and positive maxima at
$\Theta\sim 60^\circ$ and $\Theta\sim 120^\circ$, respectively.
 Moreover, the maximum value of 
the GM force component $(f_{_{\rm GM}})_\Theta$ is larger than that of
$(f_{_{\rm GM}})_r$ by factors of $\simeq 10.5$ 
and $16.4$ at $r_{\rm mb}$
and $r_{\rm ph}$, respectively, 
and the  strength of $(f_{_{\rm GM}})_\Theta$ 
decreases
little with increasing $r$ [cf.~Fig.~2(a) and~2(b)].  
 Figure~2  will be discussed further in the following section.


%
\section{Discussion: The GM 
Field and the Penrose
 Processes} 
\label{sec:3}

In this section, I use the results 
of the four-momenta of the Penrose 
scattered particles, found in [5], 
and presented again in Eqs.~(\ref{eq:15}) through (\ref{eq:18})  
and~(\ref{eq:24}) through (\ref{eq:27}) showing specific
details, to see 
what internal role the GM force of 
Eqs.~(\ref{eq:40}) through (\ref{eq:42})
 has in determining the 
spacial distributions  of the Penrose 
scattered escaping particles.
In the following,
one must bear in mind
that the GM force has already been accounted for in these Penrose
processes (as done in the model calculations of [5]) 
through the Kerr metric tensor of Eq.~(\ref{eq:1}), whose
function
is to describe the geodesics of the  particles, on which
the four-momentum is defined. 
That is, just as we do not have to calculate 
separately the gravitational
force on the scattered particles [Eq.~(\ref{eq:35})],  
we do not have to calculate separately
the GM force on the scattered particles [Eq.~(\ref{eq:37})], 
because both of these 
gravitational forces are inherently tied to the overall scattering
process through the Kerr metric.  The GM  force, however,  is
given here separately only to show how it relates to the final
four-momentum vectors of the scattered particles.
Therefore, when looking at this force separately, in
some instances, the precise relationship of the GM force with a
distribution of scattered particles may not be clear.
With this in mind we now proceed.

\subsection{The Gravitomagnetic Field and PCS}
\label{sec:3a}
 
Beginning with the PCS, displayed in Figs.~3$-$7 
are scatter plots of 
the momentum components $(P_{\rm ph}^\prime)_r$ and  
$(P_{\rm ph}^\prime)_\Theta$ [see Eqs.~(\ref{eq:16}) 
and~(\ref{eq:17})],
and the corresponding polar 
angles $\Theta_{\rm ph}^\prime$ [to be 
compared with $\Theta$ in Eqs.~(\ref{eq:40}) 
through~(\ref{eq:42})] of the
escaping PCS photons. The polar angle (defined here as the angle 
of escape [3,4]) is
given generally for a particle  of type $i$ by [25]
\begin{equation}
\Theta_i={\pi\over 2}\mp
\arccos{\left[ {-T+\sqrt{T^2-4SU}\over 2S} \right] }^{1/2},
\label{eq:43}
\end{equation}
where
\begin{eqnarray*}
S&\equiv& a^2(\mu_o^2-E^2), \\
T&\equiv& Q +a^2(E^2-\mu_o^2)+L^2, \\
U&\equiv& -L^2; 
\end{eqnarray*}
 the negative and positive signs are for particles traveling
above or below the equatorial plane, respectively.   
The polar angle of Eq.~(\ref{eq:43}) is derived 
from Eq.~(\ref{eq:12}) by letting
$P_\Theta\longrightarrow 0$.
(Note, such scatter plots as displayed in the 
figures presented
here result from the application   
of the Monte Carlo method, which yields individual scattering
angles and four-momentum components for the scattered particles.)

Upon comparing Figs.~2(a), 3, and~6, we observe the following. 
For  PCS at $r=r_{\rm mb}$, between radially infalling 
photons ($E_{\rm ph}\leq 1$~MeV) and tangentially equatorial
orbiting target electrons
 [Figs.~3(a)$-$3(e)], the spectrum appears
symmetric in $(P_{\rm ph}^\prime)_r$ for the lowest incident photon
 energy used: 
$E_{\rm ph}=0.511$~keV.   What role the GM field plays  in giving
 rise to this
symmetric distribution  is unclear [cf. Fig.~3(a)], since such a 
distribution is expected in general.  
In Figs.~3(b)$-$3(e), for the higher energy incident 
photons, it appears that GM force 
component $(f_{_{\rm GM}})_r$ of Fig.~2(a) has little effect on the
distribution of escaping  photons.  
It seems that the distribution of the scattered photons in the 
radial direction is, in these cases,  dominated by the 
momentum forces involved in the scattering process itself. That is, 
as $E_{\rm ph} $ is allowed
to increase, while $E_e$ is held constant at $0.539$~MeV,
 the preferred direction of the
scattered photon is radially inward---such
behavior is to be expected.  
In addition, even though some of photons
are scattered in the positive 
radial direction, the scattering behavior
is inconsistent with what would be expected 
if $(f_{_{\rm GM}})_r$ were
important, namely, one would expect for the particles with 
$\Theta_{\rm ph}^\prime\sim \pi/2$, and large
$(P_{\rm ph}^{\prime})_\Phi$,
and thus large $E_{\rm ph}^\prime$,
 to have positive radial momenta [cf.~
Figs.~2(a),  3(b)$-$3(e), 6(b)$-$6(e), 
and Eq.~(\ref{eq:40})].

In 
the case of PCS by  nonequatorially confined target electrons 
[Figs.~3(f)$-$3(i),
with $6.145\, M^2m_e^2\leq Q_e\leq 614.5\,  M^2m_e^2$]---i.e.,
as the energy-momentum of the target particle is allowed to 
increase, we 
find that  
the number of photons with 
outward radial momenta increases with increasing $E_{\rm ph}^\prime$ 
[implying increasing $(P_{\rm ph}^\prime)_\Phi$].  
When one compares the higher energy photons and their
polar angles of escape,
displayed in Figs.~6(f)$-$6(i), 
with Fig.~2(a) and eq.~(40), it 
appears that $(f_{_{\rm GM}})_r$ may 
have some importance in the behavior causing the number
of photons scattered with positive  radial momenta to increase 
[cf. Figs.~3(f)$-$3(i)].  
Such behavior could be due, in part, to the fact that 
$(f_{_{\rm GM}})_r$ 
is large
on the particles with large $(P_{\rm ph}^{\prime})_\Phi$ 
(i.e., large
$E_{\rm ph}^\prime$) and with $\Theta_{\rm ph}^\prime$ 
near $\sim \pi/2$; 
this is consistent with what we find in the cases of 
Figs.~3(f)$-$3(i) and corresponding angles 6(f)$-$6(i).   
The other part responsible for the behavior  
 could be that, as $E_e$ become $\gg$
 $E_{\rm ph}$, with increasing $Q_e$, backward scattering 
[$(P_{\rm ph}^{\prime})_r>0$] becomes more effective:
cf.~3(d) to~3(f) and~3(g), cf.~3(c) to~3(h) and~3(i), 
showing progression towards symmetry  as
$E_e$ becomes $\gg$ $E_{\rm ph}$. 
We will return to this discussion 
 after the introduction of Eq.~(\ref{eq:47}).
 
Before proceeding, note the following:  
the nonequatorially confined  
(spherical-like orbiting) target electrons discussed here, 
and target photons 
discussed in the next section, for a given distribution,  
can have either $P_\Theta<0$ or $P_\Theta>0$, as the particle 
passes through the equatorial plane going above or below, 
respectively.
In the figures shown here, one-half of the distribution of 2000 targets 
are given the 
negative $P_\Theta$ value and other half are given the corresponding 
positive $P_\Theta$ value (this was not done in
[5]).   Such a distribution yields  what would likely be
 seen by an observer at infinity.  Also, note that, 
consideration has been
given  only to collisions between
infalling incident particles and bound orbiting target particles, 
thus, assuming that collisions
between neighboring  target particles are negligible.

Continuing,  now the discussion centers on the various 
distributions of PCS photons  as they escape
in the positive and negative ${\bf \hat e_\Theta}$ directions.
Upon comparing Figs.~2(b) and  4$-$7, we observe 
that the polar angles $\Theta_{\rm ph}^\prime$ of
the PCS photons, after scattering off
equatorial target electrons [Figs.~6(a)$-$6(e)] 
and nonequatorially
confined target electrons [Figs.~6(f)$-$6(i) and 
7(a)$-$7(d)], 
are consistent with the  GM force component 
$(f_{_{\rm GM}})_\Theta$ [Eq.~(\ref{eq:41})] 
acting on the scattered particles.
The effects can be seen in the polar coordinate 
momenta $(P_{\rm ph}^\prime)_\Theta$, displayed in 
Figs.~4 and~5---corresponding to cases of Figs.~6 and~7, 
respectively.  
Looking closely at the distributions 
resulting from the equatorial target electrons, 
the PCS photons go
from the general expected symmetric distribution,  for the 
lowest monochromatic incident photon 
energy used: $E_{\rm ph}=5.11\times 10^{-4}$~MeV, to being strongly 
asymmetrical or one-sided, for the highest energy used:
$E_{\rm ph}=1$~MeV, with $E_e\simeq 0.539$~MeV being the same
 for all the cases;   
cf. Figs.~4(a)$-$4(e).   
Overall, we find that, as the energy of 
the incident photon $E_{\rm ph}$ is increased, more and more 
of the photons are scattered with $(P_{\rm ph}^\prime)_r<0$,
$(P_{\rm ph}^\prime)_\Theta>0$  [i.e., with $\Theta_{\rm ph}^
{\prime}>90^\circ$; cf. Figs.~3(a)$-$3(e), 4(a)$-$4(e),
 and~6(a)$-$6(e)].    
Apparent reasons for such 
distributions are explained in the following.

Recall that the magnitude of the GM potential is
$\mid \vec\beta_{_{\rm GM}}\mid=\omega$ (see Section~\ref{sec:2d}).
By examining the angle of
Eq.~(\ref{eq:21}),
we can see how this GM potential, producing 
the GM tensor field  [see Eqs.~(\ref{eq:37}) and~(\ref{eq:38})], 
causing the frame dragging,
may alter the incoming angle
$\theta_{\rm ph}^R$ of the incident
photon, and subsequently the outgoing angles,
$\theta_{\rm ph}^{R'}$ and  $\phi_{\rm ph}^{R'}$,
of the scattered photon, given
by Eq.~(\ref{eq:22}) and Eq.~(3.39) of [5]:
\begin{equation}
\phi_{\rm ph}^{R'}=\phi_{\rm ph}^R-\arccos\left({{\cos\delta^R
           -\cos\theta_{\rm ph}^R\cos\theta_{\rm ph}^{R'}}\over
       {\sin\theta_{\rm ph}^R\sin\theta_{\rm ph}^{R'}}}\right),
\label{eq:44}
\end{equation}  
respectively, where $\phi_{\rm ph}^{R'}$, the important angle here,
 gives the 
direction of the scattered photon, above and  below the equatorial 
plane.  [Note, the angles
$\theta_{\rm ph}^{R'}$ and  $\phi_{\rm ph}^{R'}$ define the 
space momentum components as described by a unit four-vector 
tangent to the 
coordinate lines at the event $(r,\Theta,\Phi,t)$, in the ERF [5];
this particular LNRF coordinate system is centered on the electron 
with the
pole in the direction of positive $\bf \hat e_\Phi$.]
So, in essence,  what we are finding is that, the GM force
close to event horizon,
whose presence is seen by the appearance of the frame dragging 
velocity $\omega$ 
[cf. Eqs.~(\ref{eq:22}) and~(\ref{eq:44})], 
 exerts a force on the 
moving particles, causing them to behave in various 
general relativistic ways (uncharacteristic of Newtonian physics), 
as discussed further below.   

Specifically, for the cases of Figs.~4(a)$-$4(e), 
and corresponding
Figs.~6(a)$-$6(e), Eq.~(\ref{eq:21}) shows that 
$\theta_{\rm ph}^{R}$ will not change for PCS off monochromatic
distributions of 2000 target electrons, at $r_{\rm mb}$, if 
$E_e$ ($\simeq 0.539$~MeV)  is
held constant for the different cases while $E_{\rm ph}$ is allowed
to increase, increasing from 
$5.11\times 10^{-4}$~MeV to $1$~MeV.   But, increasing $E_{\rm ph}$ 
will, however,
change the scattering angle $\delta^R$,  subsequently, still, altering
the scattering angles $\theta_{\rm ph}^{R'}$ and 
$\phi_{\rm ph}^{R'}$ [cf. Eqs.~~(\ref{eq:22}) and~(\ref{eq:44})].
So, how does $\delta^R$ relate to the frame 
dragging velocity
$\omega$?
The angle $\delta^R$ is determined, as
stated earlier, from the Klein-Nishina cross section: the effective 
area of an electron producing a scattering event in which a photon 
is emitted at particular polar and azimuthal angles, $\delta^R$ and
$\alpha^R$, respectively, of a spacetime coordinate 
system centered on 
the electron, with the pole in the direction of $-\bf \hat e_r$ 
(i.e., of the initial photon direction). 
The angle $\delta^R$ depends on the Lorentz factor $\gamma_e$ 
of the target electron
and the initial incident photon energy $E_{\rm ph}$ [see 
Eq.~(\ref{eq:20})].  Equation~(\ref{eq:19}) reveals the direct
relationship between $\gamma_e$ and $\omega$, and, thus, $\delta^R$. 

Investigating further, 
the angle $\phi_{\rm ph}^{R'}$, 
through the space momentum
vector of the scattered photon in the ERF [5]:
\begin{eqnarray} 
(p_{\rm ph}^{R'})_r&=&p_{\rm ph}^{R'}\sin\theta_{\rm ph}^{R'}
                     \cos\phi_{\rm ph}^{R'}, \nonumber \\
(p_{\rm ph}^{R'})_\Theta&=&p_{\rm ph}^{R'}
      \sin\theta_{\rm ph}^{R'}\sin\phi_{\rm ph}^{R'},
\nonumber   \\     
(p_{\rm ph}^{R'})_\Phi&=&p_{\rm ph}^{R'}\cos\theta_{\rm ph}^{R'},
\label{eq:45}
\end{eqnarray}
gives, in general,  in the four quadrants relative to the 
equatorial plane, that
\begin{eqnarray}
{\rm I:}~~0&&\,\leq\phi_{\rm ph}^{R'}\leq 90^\circ,\qquad\qquad\qquad
{\rm II:}~~\,90^\circ\leq\phi_{\rm ph}^{R'}\leq 180^\circ,
\nonumber\\
&&(p_{\rm ph}^{R'})_r\geq 0, \quad\qquad\qquad\qquad\qquad\qquad
(p_{\rm ph}^{R'})_r\leq 0,\nonumber\\
&&(p_{\rm ph}^{R'})_\Theta\geq 0, \quad\qquad\qquad\qquad\qquad\qquad
(p_{\rm ph}^{R'})_\Theta\geq 0,\nonumber\\
\quad 90^\circ&&\leq \Theta_{\rm ph}^\prime\leq 180^\circ;\qquad\qquad
\qquad\qquad
90^\circ\leq\Theta_{\rm ph}^\prime\leq 180^\circ;
\nonumber\\
&&~~\nonumber \\ 
{\rm III:}~~180^\circ&&\leq\phi_{\rm ph}^{R'}\leq 270^\circ,
\qquad\qquad\qquad
{\rm IV:}~~270^\circ\leq \phi_{\rm ph}^{R'}\leq 360^\circ,\nonumber\\
&&(p_{\rm ph}^{R'})_r\leq 0, \qquad\qquad\qquad\qquad\qquad\qquad
(p_{\rm ph}^{R'})_r\geq 0,\nonumber\\
&&(p_{\rm ph}^{R'})_\Theta\leq 0, \qquad\qquad\qquad\qquad\qquad\qquad
(p_{\rm ph}^{R'})_\Theta\leq 0,\nonumber\\
\quad 0&&\leq\Theta_{\rm ph}^\prime\leq 90^\circ;\qquad\qquad
\qquad\qquad\qquad\,
0\leq\Theta_{\rm ph}^\prime\leq 90^\circ.
\label{eq:46}
\end{eqnarray}
We will use the quadrants of Eq.(\ref{eq:46}), in the
discussion  below, in an effort to further
understand the distribution of PCS photons. 
(Note, the Lorentz spacetime transformations for eq.(45),
from the ERF to
a general LNRF observer, are given in [5] [eq.~(3.46)]).                  

Upon comparing Eq.~(\ref{eq:46}) and Figs.~4 and~6, we find 
the following.  In Fig.~4(a) [see also Fig.~3(a)], the photons
are scattered symmetrically into the four quadrants.  Increasing
 $E_{\rm ph}$ causes the highest energy photons
to be scattered below the equatorial
plane, and the appearance of a void in the lowest energy regime;
cf. ~Figs.~4(a)$-$4(e). These figures show how asymmetry 
in the preferred positive ${\bf \hat e_\Theta}$ direction, of
$(P_{\rm ph}^\prime)_\Theta$, increases as $E_{\rm ph}$  
is increased.
The highest energy photons  
are progressively scattered 
into quadrant II, consistent with $(f_{_{\rm GM}})_\Theta$ [see 
Eq.~(\ref{eq:41}) and Fig.~2(b)]
acting below the equatorial plane on these scattered particles, 
proportional to increasing
$(P_{\rm ph}^\prime)_\Phi$ (i.e., increasing $E_{\rm ph}^\prime$),
 with $90^\circ\, 
{<\atop\sim}\,\Theta_
{\rm ph}^\prime\,{<\atop\sim}\, 180^\circ$.
   On the other hand, the lowest energy
photons are progressively scattered into quadrant III,
consistent with $(f_{_{\rm GM}})_\Theta$
acting above the equatorial plane, producing a smaller GM force on
these particles, and thus a smaller magnitude for 
$(P_{\rm ph}^\prime)_\Theta$,  with $0\,
{<\atop\sim}\,\Theta_
{\rm ph}^\prime\,{<\atop\sim}\, 90^\circ$ [cf. Figs.~6(a)$-$6(e)].
The most probable explanation for these distributions as related 
directly to the GM field, although not apparent from 
Eqs.~(\ref{eq:40}) through~(\ref{eq:42})---which give the GM 
force as measured by a LNRF observer, will be given below.

So, overall, the equatorial 
target electrons scatter the photons preferentially 
in the direction of positive 
${\bf \hat e_\Theta}$, as the energy-momentum of the incident
photon is increased.   
The asymmetry seen in the scattered 
particle distributions described above, at these relatively 
low incident and target particle energies, while 
$E_e$ ($\simeq 0.539$~MeV) is held constant, 
can be viewed 
as the ``signature'' of the GM field acting on the
infalling [$(P_{\rm ph})_r<0$] incident photons, and
the scattered photons, 
while increasing $E_{\rm ph}$ 
(i.e., increasing $\vert (P_{\rm ph})_r\vert$).  
The resulting effects can be seen in the momentum 
component of 
Eq.~(\ref{eq:17}), plotted in Figs.~4(a)$-$4(e), 
in which the angles 
contained therein are altered by certain combinations
of $\omega$ and the energy-momenta of the incident
and scattered particle energies.    
The ratio of the number of escaping photons with
$(P_{\rm ph}^\prime)_\Theta>0$ to those with
$(P_{\rm ph}^\prime)_\Theta<0$ ($\equiv \epsilon_i$,
for particle of type $i$)  range from
$\epsilon_{\rm ph}\simeq 1.07$, $\simeq 1.46$, $\simeq 2.42$, $
\simeq 3.42$, up to
$\simeq 4.99$, for Figs.~4(a)$-$4(e),
respectively. 
Thus, as stated earlier,
it appears that reflection symmetry of the Kerr metric,
above and below the equatorial plane, is broken due to the GM force
field.
  
Note, in general, since $(P_{\rm ph}^\prime)_\Phi$ 
is positive for all of the  escaping Penrose scattered
particles, the antisymmetric component $(f_{_{\rm GM}})_\Theta$ 
[see Eq.~(\ref{eq:41}) and Fig.~2(b)]
will exert a force in the negative ${\bf \hat 
e_\Theta}$ direction for 
$\Theta_i^\prime<90^\circ$ and in the positive ${\bf \hat 
e_\Theta} $ direction for $\Theta_i^\prime>90^\circ$;
such could aid in  the acceleration of the 
jets of AGNs.  Namely, this GM force component has  the 
effect of 
blueshifting the escaping photons to higher energies,
while increasing the Lorentz factor of the 
escaping  electrons.

Now we will consider the PCS by nonequatorially confined 
target electrons,  with the 
initial monochromatic photon energy held constant at 
$E_{\rm ph}=5.11\times 10^{-4}$~MeV, while $E_e$ is
allowed to increase from the equatorially confined value
$E_e\simeq 0.539$~MeV, $Q_e=0$  [Fig.~4(a)] up to 
$E_e\simeq 5.927$~MeV, 
$Q_e=154.5\,  M^2m_e^2$,
for the different
cases
[cf. $(P_{\rm ph}^\prime)_\Theta$ and corresponding
angles of escape $\Theta_{\rm ph}^\prime$ of Figs.~5 and~7, 
respectively].
The asymmetry proposed to be due to the GM force (indicated by 
the prominent void) is quite apparent 
in the different cases [cf.~Figs.~5(a)$-$5(c)], 
i.e.,  before, it seems, momentum forces 
inherent to the scattering processes, because of 
increasing $Q_e$, and  the final effects of the GM force acting 
on the scattered particles,  dominate over the initial effects of 
the GM force, thus, 
 causing  symmetry to  reappear,
as seen in Fig.~5(d).    For the nonequatorially 
confined target electrons,
I did not derive expressions as done for the equatorial
targets [Eqs.~(\ref{eq:15}) through~(\ref{eq:22})], showing 
the direct relationship between the momentum components
and the angles producing the asymmetry.  Nevertheless, 
it appears that the 
asymmetry is caused, in general,  by the same 
mechanism, i.e., the GM force alters the incoming and outgoing
angles of  Eqs.~(\ref{eq:22}) and~(\ref{eq:44}), through
the presence of $\omega$. 
To further compare, 
the ratio of the number of escaping photons with 
$(P_{\rm ph}^\prime)_\Theta>0$ to those with 
$(P_{\rm ph}^\prime)_\Theta<0$ 
 range from 
$\epsilon_{\rm ph}\simeq 1.07$ for Fig.~4(a): $Q_e=0$ to 
$\epsilon_{\rm ph}\simeq 1.54$, $\simeq 2.13$, $\simeq 1.29$,  
$\simeq 1.02$, for Figs.~5(a)$-$5(d): $Q_e\neq 0$,
respectively.   The specific cause of the behavior for
 these scattered photon distributions,  varying
 from symmetric to asymmetric
[scattering preferentially into quadrants I and II; see
Eq.~(\ref{eq:46})],
 then back to symmetric (scattering equally into the four 
quadrants), is not readily
accessible from the GM force components [Eqs.~(\ref{eq:40})
through~(\ref{eq:42})], but, again, the presence
of the frame velocity in the scattering angles
[Eqs.~(\ref{eq:22}) and~(\ref{eq:44})] assures us that the GM
force field 
has a role in determining the trajectories of the photons. We
will return to these distributions 
 below, where we will find that 
the asymmetry seen in the scattered
particle distributions, at the relatively
low incident and target particle energies
[Figs.~5(a)$-$5(c)], while
$E_{\rm ph}$ ($=5.11\times 10^{-4}$~MeV) is held constant,
can be viewed (as in the case of the equatorially confined target 
electrons described above) as the signature of the GM field 
acting on the
initial infalling photons, and the scattered photons.   [Note that,
the highest energy photons are scattered with polar escape angles
$\Theta_{\rm ph}^\prime$ approximately equal to the polar angle
indicating the maximum or minimum latitudinal angle of the 
spherical-like
orbiting target electrons ($\simeq 29.5^\circ$ relative to 
the equatorial
plane), given by 
Eq.~(\ref{eq:43}); cf.~Fig.~7(d).]         

Deviating briefly with a discussion to help us to 
understand better the above scattered particle distributions
and those to follow:  we do not yet 
have  analytic expressions  for the force components 
[like  Eqs.~(\ref{eq:40}) through~(\ref{eq:42})] 
describing the behavior of the GM 
 field on a
test particle  near the event horizon as measured 
by a local BLF (LBLF) observer.
The reason for this is because of the complexity in deriving
an analytic expression for the
GM tensor field $\tensor{\bf H}$ inside the ergosphere in the LBLF, 
i.e., due to the severe
dragging of local inertial frames, in which the time coordinate
basis vector $(\partial /\partial t)$ of the Kerr metric 
changes from timelike to spacelike [7].  Algebraically, this
complexity arises in the partial derivatives of the
Kerr metric components, contained in the  nonzero 
Christoffel symbols
$\Gamma_{i j}^k$ (sometimes referred to as the affine 
connections),  when evaluating the covariant derivatives in the
general expression for $H_{ij}$ [cf.~Eq.(\ref{eq:37})]. 
Therefore,  until an exact expression for $\tensor{\bf H}$ can be
found, we can only approximate its physical
nature based on 
the information 
we have at hand,  and what these Penrose 
scattering processes
are telling us, about the behavior of the GM field
close to the event horizon.  
Now, we know
that the GM field at radii $\gg$ than the event horizon 
($r=r_+$)
appears as a dipolar-like field as measured by a BLF
observer.  This
dipolar-like field will however
 be distorted near the event horizon 
as measured by this distant observer, due to large velocities and 
strong gravity, but not fully
destroyed [10].  It is possible that the preference of the particle
being scattered into the positive ${\bf \hat e_\Theta} $ direction
 is a display 
of this distorted 
dipolar-like field.  
The most probable explanation for the preference in the 
positive ${\bf \hat e_\Theta}$ direction is that, near the
event horizon in the 
BLF, the GM force field lines are frame dragged into the direction
of the rotating KBH, i.e.,  positive 
azimuthal (${\bf \hat e_\Phi}$) direction, which  
 results in $\tensor{\bf H}$, the GM tensor field,
acquiring additional cross terms: $H_{r\Theta}$ and $H_{\Theta r}$
[cf.~Eq.~(\ref{eq:38})].   This is supported by the
acquired nonzero $\Gamma_{i j}^k$'s  in the covariant 
derivatives---appearing in
the  component notation for $\tensor{\bf H}$ [see Eq.~(\ref{eq:37})], 
which in general
gives rise to 
nonzero $H_{r \Theta}$ 
and $H_{\Theta r}$ components.  In this case, the GM vector field,
say $\bf H$, will
acquire an azimuthal component $H^\Phi$, such that  
${\bf H}=H^r{\bf \hat e_r}+H^\Theta{\bf \hat e_\Theta}+
H^\Phi{\bf \hat e_\Phi}$ [cf.~Eq.~(\ref{eq:39})]. Then, by the
analogy of the approximation in the asymptotic rest (or
nonrotating) BLF: 
${\bf F}_{_{\rm GM}}\simeq
{\bf P}\times{\bf H}$, for $r\gg r_+$, and using Eq.~(\ref{eq:37}),
we find 
that the GM force on a test particle
as measured by a LBLF (inertial framed dragged)  observer, for
$r\sim r_+$, can be expressed by the following:
\begin{eqnarray}
{\bf F}_{_{\rm GM}}&=&{\bf
\tensor{\bf H} \cdot p} \nonumber \\
&=&\big[(F_{_{\rm GM}})_r,(F_{_{\rm GM}})_\Theta,
(F_{_{\rm GM}})_\Phi\big] \nonumber \\
&=&(H_{r\Theta} \,p^\Theta+H_{r\Phi} \,p^\Phi)
      {\bf \hat e_r}+(H_{\Theta r} \,p^r+H_{\Theta\Phi}\, p^\Phi)
{\bf \hat e_\Theta}
        +(H_{\Phi r} \,p^r+H_{\Phi\Theta}\, p^\Theta)
           {\bf \hat e_\Phi} \nonumber \\ 
&\propto& 
     (H^\Phi P_\Theta-\tilde H^\Theta P_\Phi)
      {\bf \hat e_r}+(H^r P_\Phi-\tilde H^\Phi P_r)
       {\bf \hat e_\Theta}
        +(H^\Theta P_r-\tilde H^r P_\Theta){\bf \hat e_\Phi}.
\label{eq:47}
\end{eqnarray}
Note, the $\propto$ indicates that we are assuming 
(1) the space
metric tensor $g^{j k}$, multiplying
the covariant momentum components has been defined by a local
orthonormal (Lorentz) tetrad $h_\alpha^\mu$ so that, as usual, 
$g^{\mu\nu}=h_\alpha^\mu h_\beta^\nu \eta^{\alpha \beta}$ with 
$\eta^{\alpha\beta}$ being the Minkowski metric components; 
and (2) the space momentum transformations from the LBLF to the 
global BLF
 are standard [7], i.e., ``somewhat'' similar 
to transformations from the LNRF to the global BLF [5].  
It is conceived that 
the above assumptions are at least valid for our analyzing purposes.  
Again, I remind 
the reader that the GM force presented in Eq.~(\ref{eq:47}) is derived
here only to show how the particles might be affected by the GM field, but 
these effects, shown in the resulting distributions of the escaping
particles as measured by a BLF observer, are intrinsically incorporated 
into the calculations through physical processes occurring in the Kerr metric 
spacetime geometry.
The components of the
vectors ${\bf H}$ and ${\bf\tilde H}$ in Eq.~(\ref{eq:47}) 
are defined by the 
 components of the  
GM tensor:
\begin{eqnarray}
H^r\equiv H_{\Theta\Phi},&~~~~~~~&
        \tilde H^r\equiv - H_{\Phi\Theta}, \nonumber \\ 
H^\Theta\equiv H_{\Phi r},&~~~~~~~&
         \tilde H^\Theta\equiv -H_{r\Phi}, \nonumber \\
H^\Phi\equiv H_{r \Theta};&~~~~~~~&
          \tilde H^\Phi\equiv -H_{\Theta r} 
\label{eq:48}
\end{eqnarray}  
 [cf.~Eqs.~(\ref{eq:38}), (\ref{eq:39}), 
and~(\ref{eq:40}) through~(\ref{eq:42})]. 
If the GM tensor is antisymmetric [see Eq.~(\ref{eq:39})],
then $H^r= \tilde H^r$, $H^\Theta=\tilde H^\Theta$, 
$H^\Phi=\tilde H^\Phi$.

From a geometrical analysis of the characteristics of the 
antisymmetric GM tensor components $\tensor{\bf H}$ in 
the asymptotic rest BLF (ABLF)
[see Eq.~(\ref{eq:39})] and 
 the  components of 
$\tensor{\bf H}$ in the LNRF in which the frame dragging
is ``taken out'' [see Eq.~(\ref{eq:38})], and the probable 
condition in the LBLF, that the GM tensor field lines are 
frame dragged 
in the azimuthal direction, we can deduce that the GM
tensor in the LBLF will be somewhat similar to that in the LNRF,
in respect to the dipolar-like characteristics, with 
differences, 
in general, depending on the frame dragging velocity $\omega$
[cf.~Eq.~(\ref{eq:37})].   
The tensor
components of Eq.~(\ref{eq:38}), when evaluated at $r\sim r_+$,
reveal to us the following:
\begin{eqnarray}
H^r\leq 0 ~&{\rm and}&~\tilde H^r\leq 0~{\rm for}~
\Theta\leq 90^\circ; \nonumber \\
~H^r\geq 0 ~&{\rm and}&~
\tilde H^r\geq 0~{\rm for}~\Theta\geq 90^\circ;
  \nonumber \\
H^\Theta>0 ~&{\rm for}&~ \Theta\leq 90^\circ~
{\rm and}~ \Theta>90^\circ;
        \nonumber \\
\tilde H^\Theta<0~ &{\rm for}&~ \Theta\leq 90^\circ~{\rm and}~
 \Theta>90^\circ;
        \nonumber \\
H^\Phi<0 ~&{\rm for}&~ \Theta\leq 90^\circ~
{\rm and}~ \Theta>90^\circ;
        \nonumber \\
\tilde H^\Phi>0~ &{\rm for}&~ \Theta\leq 90^\circ~{\rm and}~
 \Theta>90^\circ;
\label{eq:49}
\end{eqnarray}
cf.~Eqs.~(\ref{eq:47}) and~(\ref{eq:48}).  The additional components
included in Eq.~(\ref{eq:49}), $H^\Phi$ and $\tilde H^\Phi$, are 
found from deductive 
reasoning as follows.
We know from a general direct evaluation of the covariant derivative 
of $\vec \beta_{_{\rm GM}}$ that $H_{r\Theta}=H_{\Theta r}$.
Therefore assuming that $H_{r\Theta}=H_{\Theta r}> 0$ 
(as would be expected
because of the frame dragging), then by definition [Eq.~(\ref{eq:48})]
and the results of the Penrose scattered particle distributions, we can 
deduce the general signs of $H^\Phi$ and  $\tilde H^\Phi$ 
in, above, and below the equatorial plane. 
We will use the relations of  
Eq.~(\ref{eq:49}), as we continue in the
discussion below. 

Note, the differences between the
frames of reference considered here are the following.
An observer in the BLF measures coordinates $r$, $\Theta$, $\Phi$,
and $t$, relative to the KBH.  The BLF is an inertial frame (i.e.,
it is a freely falling coordinate system).
An observer in this frame measures spacetime events 
from two points of view: 
the asymptotic rest (or flat) inertial frame
(events occurring at $r\gg r_+$)---which I refer to as the ABLF,
 and the global inertial frame 
(events occurring at $r\sim r_+$)---which I refer 
to as the BLF or global BLF.  The measured four-momenta 
[$P_\mu=(P_r, P_\Theta, L, -E$)], describing the trajectories of the 
particles [cf.~Eqs.~(\ref{eq:12}) through~(\ref{eq:14})],  
are the same parameters for both these observers, 
although the immediate surroundings of the particles are very
different.  The BLF observer at $r\sim r_+$ 
 measures the effects of frame dragging, whereas the 
ABLF observer does not, yet both observers are located at infinity.
For example, the GM tensor field as measured by 
 the ABLF observer  
 will have 
prominent  dipolar-like features [cf.~(\ref{eq:39})], 
features that will be 
distorted as measured by the LBLF and BLF observers.
An observer in the LBLF ``sees'' this distorted field, 
whose force lines are frame dragged into the 
azimuthal direction, and measures its effects on 
moving test 
particles [cf.~Eq.~(\ref{eq:47})], but the  
ABLF observer
 cannot measure these near event horizon effects; 
it is only through the ``eyes'' of the
LBLF and BLF observers, as evident in these Penrose process distributions,
that the observer at infinity can measure these ergospheric effects.            
Now, on the other hand, the  LNRF observer   
is a local inertial frame rotating 
with the geometry [7],   
experiencing in a sense  no frame dragging. 
 This spacetime coordinate system is at rest relative to the
frame dragging, and revolves with velocity $\omega$ relative 
to the observer at infinity.  Physical processes described in 
this local Lorentz frame are ``simpler'' because the frame
dragging is canceled as much as possible.  
The observer in the LNRF
measures the force that the  local gravitationally
distorted dipolar-like GM tensor 
field exerts on a  moving test particle, however, free of frame
dragging effects.  Therefore,  the LNRF has its limitations in
 explaining 
 the local effects of the  frame dragged GM force
field; for this reason, the LBLF 
is needed.  

Equation~(\ref{eq:47})
shows that the GM force on a test particle in the LBLF and the BLF
(wherein we are assuming the most probable condition)
will acquire additional  terms in the radial and
polar components, different from that in the  ABLF 
and the LNRF.  [Note, in the LNRF, Eqs.~(\ref{eq:40}) 
through~(\ref{eq:42})
have the same general form as Eq.~(\ref{eq:47}), 
however, with $H^\Phi=\tilde H^\Phi=0$.] 
These additional terms involving $H^\Phi$ and $\tilde H^\Phi$, 
as we shall see, 
 can explain the asymmetries that we observe in
the Penrose scattered particle distributions presented here.  
For a preliminary example,
if $P_r$ is less than zero, and allowed to increase in magnitude
for the different cases considered,
 as for the initial radially infalling photon
[$(P_{\rm ph})_r<0$, $(P_{\rm ph})_\Theta=(P_{\rm ph})_\Phi=0$]
displayed in Figs.~3(a)$-$3(e); 4(a)$-$4(e); 6(a)$-$6(e), 
 with the target particle having relatively
low energy-momentum,
 Eq.~(\ref{eq:47}) shows that $(F_{_{\rm GM}})_\Theta$ will
exert a dominant force on the photon, 
increasing in the positive ${\bf \hat e_\Theta}$ direction,
and thus, initializing the asymmetry eventually seen in the high energy 
PCS photons [cf.~Figs.~4(a)$-$4(e) and 6(a)$-$6(e)].  
Now, with this initialized asymmetry, the final
distributions are consistent with  $(F_{_{\rm GM}})_\Theta$, 
again as given by Eq.~(\ref{eq:47}),  acting directly on the 
PCS photons:  In addition to this force component being positive 
for photons
scattered with ($P_{\rm ph}^\prime)_r<0$ and  $H^r\geq 0$
[i.e., $\Theta_{\rm ph}^\prime\geq 90^\circ$; cf.~Eq.~(\ref{eq:49})], 
this force component  will be 
positive for photons
scattered with ($P_{\rm ph}^\prime)_r>0$ and  $H^r>0$ 
[i.e., $\Theta_{\rm ph}^\prime>90^\circ$]
provided $H^r(P_{\rm ph}^\prime)_\Phi>\tilde H^\Phi 
(P_{\rm ph}^\prime)_r $, 
as for the
 photons with large $L_{\rm ph}^\prime$ or small 
$(P_{\rm ph}^\prime)_r$---otherwise this 
force component will be negative.
 For $(P_{\rm ph}^\prime)_r<0$ and $H^r< 0$
(i.e., $\Theta_{\rm ph}^\prime< 90^\circ$), $(F_{_{\rm GM}})_\Theta$
will be $>0$   for sufficiently
small $L_{\rm ph}^\prime$ or sufficiently large 
$\vert(P_{\rm ph}^\prime)_r\vert$---otherwise $(F_{_{\rm GM}})_\Theta$
will be $<0$.
This behavior is  consistent with Figs.~4(a)$-$4(e). 
On the other hand, if the energy-momentum of the infalling incident 
photon is relatively small compared to that of the orbiting target
electron [in which, say  $(P_e)_\Phi$ is allowed to increase for different
cases, with $(P_e)_r=0$], Eq.~(\ref{eq:47}) shows that 
$(F_{_{\rm GM}})_\Theta$, exerted on the scattered photon, will 
dominate, creating or recreating symmetry in the polar direction.  
We find this  behavior to be consistent with what we 
observe in the PCS photons of 
Figs.~4(a), 5, 6(a), 7. 
More on this later.

So, in general, Eq.~(\ref{eq:47}) shows that it is the 
$\tilde H^\Phi$ component of the GM tensor
 field in the LBLF, acting on 
radial component of the 
infalling incident particles, that subsequently causes the polar angle
$\Theta_{\rm i}^\prime$, of the scattered particles of type $i$,
to be $> 90^\circ $ (i.e., $P_\Theta^\prime >0$).  
Once the preference of the scattering angle 
$\Theta_{\rm i}^\prime>90^\circ$ 
 has been established by the GM force as measured
by a LBLF observer, we find 
that the scattered particle trajectories are somewhat consistent
with the GM force  as measured by a LNRF observer 
[ Eqs.~(\ref{eq:40}) through~(\ref{eq:42})]---however, the 
trajectories cannot be fully explained in this
frame.   Therefore, in the remainder of this
discussion, we will now use  Eq.~(\ref{eq:47}), the proposed, 
most probable GM force  measured by a 
LBLF observer, which includes the use of
Eqs.~(\ref{eq:40}) through~(\ref{eq:42}) as measured by
a LNRF observer, to further show the effects that the GM field 
has on these Penrose scattering processes---remembering that 
 the BLF observer, i.e., the observer located at infinity measures 
both near event horizon events ($r\sim r_+$ $\equiv$ BLF or global BLF) and  
 faraway events ($r\gg r_+$ $\equiv$ ABLF). 

Before proceeding, we return to the cases of
Fig.~5 to shed 
some additional light on these 
distributions using
Eq.~(\ref{eq:47}). 
The radial distributions of these PCS photons remain nearly 
symmetrical for the different cases (not displayed here), 
looking 
very much like  the case of Fig.~3(a), where there is a slight
preference for the number of particles escaping with 
$(P_{\rm ph}^\prime)_r<0$.  
Using Eq.~(\ref{eq:49}),
Eq.~(\ref{eq:47})  shows in general that for $H^r\geq 0$
(i.e., $\Theta_i\ge 90^\circ$) and  $P_r<0$, 
$(F_{_{\rm GM}})_\Theta$ is $>0$. But for $P_r>0$  and  $H^r>0$
(i.e., $\Theta_i> 90^\circ$),
$(F_{_{\rm GM}})_\Theta$ is $>0$ or $< 0$ for sufficiently 
large or small azimuthal
momentum $P_\Phi$ ($= L$), respectively.  On the other 
hand, 
Eq.~(\ref{eq:47})  shows in general that for $H^r\leq 0$
(i.e., $\Theta_i\leq 90^\circ$) and  $P_r>0$,
$(F_{_{\rm GM}})_\Theta$ is $< 0$. But for $P_r<0$ and 
$H^r< 0$ (i.e., $\Theta_i< 90^\circ$), 
$(F_{_{\rm GM}})_\Theta$ is $<0$ 
or $>0$ for sufficiently
large or small azimuthal
momentum $P_\Phi$, respectively.  
Thus, the above analysis shows possible symmetry in 
the polar direction
if there is symmetry in $P_r^\prime$ of the scattered 
particles distribution.  However, if there is 
asymmetry in $P_r^\prime$,
say, the particles are scattered preferentially
with $P_r^\prime<0$, there will be asymmetry in the polar 
direction: For $P_r^\prime <0$, the  component 
$(F_{_{\rm GM}})_\Theta$ acting on the
particles with $\Theta_i^\prime\geq 90^\circ$ will be $>0$, but 
$(F_{_{\rm GM}})_\Theta$ acting on 
particles with $\Theta_i^\prime<90^\circ$ 
will be $>0$ for particles with $\vert H^r P_\Phi^\prime\vert<
\vert\tilde H^\Phi P_r^\prime\vert$  (implying small $L^\prime$, 
low $E^\prime$ or large $\vert P_r^\prime\vert$) and  $<0$ for 
particles with $\vert H^r P_\Phi^\prime\vert>
\vert\tilde H^\Phi P_r^\prime\vert$ (implying
large $L^\prime$, high $E^\prime$ or small
$\vert P_r^\prime\vert$).  This behavior of 
Eq.~(\ref{eq:47})  is sufficient to give a general explanation
for  the polar asymmetries occurring in these Penrose 
scattering processes if there
is asymmetry in the scattered particle spectral distribution of 
$P_r^\prime$ [cf.~Figs.~3(a)$-$3(e) and~4(a)$-$4(e)].
Yet, if there is symmetry in $P_r^\prime$ [or near symmetry---as 
we find for the cases
of Figs.~5], 
Eq.~(\ref{eq:47})  
shows that $(F_{_{\rm GM}})_\Theta$ acting on the scattered 
particles should produce symmetry in the polar direction.
However, because of the initial asymmetry, favoring 
the positive ${\bf \hat e_\Theta}$ direction, caused by the GM force
component $(F_{_{\rm GM}})_\Theta$ acting on the radial 
component $(P_{\rm ph})_r$ of the infalling incident photons 
[cf.~Eq.~(\ref{eq:47})], the scattered particle distributions,
at sufficiently low orbital energy-momentum for the target
particle, will not reflect the expected mirror symmetry, above and 
below the equatorial
plane---contrary to what one would expect in Newtonian physics or 
in the weak
gravitational regime ($r\gg r_+$); 
this can be seen in Figs.~5(a)$-$5(c).
That is, according to the first term in $(F_{_{\rm GM}})_\Theta$
[cf.~Eq.~(\ref{eq:47})] and the description above for small $L^\prime$
(or for the satisfying of $\vert H^r P_\Phi^\prime\vert<
\vert\tilde H^\Phi P_r^\prime\vert$), this initial asymmetry in the  
$+\bf \hat e_\Theta$ direction is amplified 
when $(F_{_{\rm GM}})_\Theta$  acts on the subsequent 
PCS photons.
But, as $Q_e$  of the nonequatorially
confined target electrons
is allowed to increase for the different cases,  i.e., increasing 
the magnitudes of 
$(P_e)_\Theta$ and $L_e$, and subsequently $(P_{\rm ph}^\prime)_\Phi$,
the first term in $(F_{_{\rm GM}})_\Theta$ 
 dominates over this initial asymmetry, 
causing the reappearance of the symmetry we observed in 
Fig.~5(d), thus, showing us the final result of  
$(F_{_{\rm GM}})_\Theta$ acting on the scattered high energy
photons, increasing in magnitude with increasing 
$L_{\rm ph}^\prime$.   So, in essence,  $(F_{_{\rm GM}})_\Theta$ 
 gives the escaping
particles with large azimuthal component momentum
a boost into opposite  polar directions, and is greater in the
$+\bf \hat e_\Theta$ direction for particles with
$(P_{\rm ph}^\prime)_r<0$ [cf.~Figs.~3(a)$-$3(e) and~4(a)$-$4(e)].

Now proceeding with the discussion after the brief deviation, 
the momenta  $(P_{\rm ph}^\prime)_\Theta$ of
Figures~4(f)$-$4(i),
from PCS of 2000 radially infalling photons (for each case)
by nonequatorially 
confined target electrons, show presence of the GM 
force acting on the particles. 
Effects of  $(F_{_{\rm GM}})_\Theta$  [of Eq.~(\ref{eq:47})]
 acting on $(P_{\rm ph})_r$ of the initial photons,
$E_{\rm ph}=0.15$~MeV [Figs.~4(f), 4(g)] and 
$E_{\rm ph}=0.03$~MeV [Figs.~4(h), 4(i)], can clearly 
be seen, by the presence of the most prominent regions
void of particles
 [cf. Figs.~4(d) and 4(e)].  This feature (viewed as the 
signature of the GM field acting on the
infalling incident photons, and the scattered photons) is somewhat
permanent at these initial energies, in the sense that 
$(F_{_{\rm GM}})_\Theta$ acting
on the PCS photons resulting from the target nonequatorially 
confined electrons, of initial 
energy $E_e\simeq 1.297$~MeV,  
$E_e\simeq 5.927$~MeV, and $E_e\simeq 11.79$~MeV 
 [Figs.~4(f)$-$4(i)], the ``symmetry producing agent,'' 
will not dominate
over the effects producing this feature [as in Fig.~5(d))].
Once the preference for $\Theta_{\rm ph}^\prime>90^\circ$
has been established by $(F_{_{\rm GM}})_\Theta$
 acting on the infalling incident photons, as described  above, 
the additional asymmetry seen in Figs.~4(f)$-$4(i)
is consistent with $(F_{_{\rm GM}})_\Theta$ acting
on the escaping photons, which are scattered predominantly with 
$(P_{\rm ph}^\prime)_r<0$ [see Figs.~3(f)$-$3(i)]. 
According to Eq.~(\ref{eq:47}), for  $(P_{\rm ph}^\prime)_r<0$, 
$(F_{_{\rm GM}})_\Theta$
will be $>0$ (or in the $+\bf \hat e_\Theta$ direction)
 for all the 
photons scattered
with $\Theta_{\rm ph}^\prime\geq 90^\circ$, but those with
$\Theta_{\rm ph}^\prime< 90^\circ$,  $(F_{_{\rm GM}})_\Theta$
will be $>0$ if 
$\vert H^r L_{\rm ph}^\prime\vert<\vert\tilde H^{\Phi}(P_{\rm ph}
^\prime)_r\vert$, as for the lowest energy PCS photons; otherwise
$(F_{_{\rm GM}})_\Theta$ will be $<0$, i.e.,
for photons scattered with sufficiently larger energies.  
This is the behavior we
observe in Figs.~4(f)$-$4(i) [cf. also 
corresponding escape angles of Figs.~6(f)$-$6(i), respectively].  
For example, notice how $(F_{_{\rm GM}})_\Theta$ 
acting on the PCS photons resulting from the nonequatorially 
confined target electrons (the symmetry producing agent)
of increasing orbital energy-momentum, for the different cases,  
 causes more and more high energy PCS photons to 
be emitted in the negative $\bf \hat e_\Theta$ direction,
and, thus, eventually causes symmetry to somewhat
reappear  except for the  GM signature
[cf. Figs.~4(h) and 4(i) to~5(d)].   
Moreover, according to Eq.~(\ref{eq:47}), as $E_e$ increases
and more and more of the photons are scattered with
$(P_{\rm ph}^\prime)_r>0$, for  $\Theta_{\rm ph}^\prime\leq 90^\circ$,
$(F_{_{\rm GM}})_\Theta$ will be $<0$, assisting in reestablishing the
symmetry [cf. Figs.~3(f)$-$3(i)].
Note, the ratio of the number of escaping photons with
$(P_{\rm ph}^\prime)_\Theta>0$ to those with
$(P_{\rm ph}^\prime)_\Theta<0$, $\epsilon_{\rm ph}$, range from
$\simeq 2.41$ for Fig.~4(f) to $\simeq 1.34$, 
$\simeq  1.16$, and
$\simeq  1.08$ for Figs.~4(g)$-$4(i),
respectively.  

Astrophysically, these results mean that an observer at infinity
will observe ``photon jets,'' collimated concentric the polar axis,
ranging from symmetric to asymmetric,
depending on the accretion disk properties: which populate
the target particle orbits and supply the infalling incident
photons.  So, overall, in summary, it appears
that the angle  $\phi_{\rm ph}^{R'}$, measured perpendicular to 
the equatorial
plane, in the ERF (a particular
LNRF observer), given by Eq.~(\ref{eq:44}), is altered by 
the presence of $\omega$, the frame dragging 
velocity, such that certain
combinations of increasing
$E_{\rm ph}$ and/or $E_e$ (i.e., increasing the 
energy-momenta of the
incident and target particles) cause the scattered particle
distributions to vary from symmetric  to asymmetric, with the
 preference
in the direction of positive $\bf \hat e_\Theta$, then back to
very nearly symmetric. 
The behavior of these distributions is in some degree 
consistent with
the GM force  as measured by a LNRF observer 
[Eqs.~(\ref{eq:40}) through~(\ref{eq:42}) ], 
but this behavior cannot  be fully explained in this frame.  
The most probable explanation, that near the event horizon
the GM field lines are frame dragged in the $+\bf \hat e_\Phi$
direction  
as measured by a 
local frame dragged inertial observer
[leading to the derivation of the general expression
Eq.~(\ref{eq:47})  for the GM force as measured locally
(at $r\sim r_+$) by a
BLF observer],  
can indeed  explain this behavior.  
We shall find this to be true also for the PPP discussed below.

\subsection{The Gravitomagnetic Field and PPP ($\gamma
\gamma\longrightarrow e^-e+$)}
\label{sec:3b}

Now considering the PPP ($\gamma \gamma \longrightarrow e^- e^+$\/),
displayed in Figs.~8 and~9 are
scatter plots of
$(P_{\mp})_\Theta$ [Eq.~(\ref{eq:26})] and the corresponding polar
angles $\Theta_\mp$ [Eq.~(\ref{eq:43})], respectively, 
of the escaping 
$e^-e^+$ pairs for various cases, after 2000
PPP  events at the photon orbit for each case. 
In general, the force component $(F_{_{\rm GM}})_\Theta$ of
Eq.~(\ref{eq:47}), using Eq.~(\ref{eq:49}),
reveals to us that infalling particles ($P_r<0$) and outgoing particles
($P_r>0$) moving along  the equatorial plane ($\Theta_i =90^\circ$) will 
experience a force $(F_{_{\rm GM}})_\Theta>0$ and 
$(F_{_{\rm GM}})_\Theta<0$, respectively.  The GM force component 
$(F_{_{\rm GM}})_\Theta>0$ acting on the infalling incident
photon of energy $E_{\gamma 1}=0.03$~MeV is expected be 
somewhat significant in
initializing the asymmetry, above and below the equatorial
plane, favoring the positive $\bf \hat e_\Theta$ direction
($\Theta_i>90^\circ$), eventually seen in the scattered
particle distributions, in the relatively low energy regime,
 as indicated in Fig.~4(c).
However, this asymmetry is not apparent in the case of Figs.~8(a)
and~9(a).  In this case it appears that, this initialized asymmetry
is balanced (or canceled) by $(F_{_{\rm GM}})_\Theta$ acting on the
outgoing $e^-e^+$ pairs---at these relatively 
low incident
and target particle energies ($E_{\gamma 2 }\simeq 13.54$~MeV), 
for $\Theta_\mp\leq 90^\circ$, and if the condition is satisfied that 
 $\vert H^rL_\mp\vert<\vert \tilde H^\Phi (P_\mp)_r\vert$ 
for $\Theta_\mp>90^\circ$,  
$(F_{_{\rm GM}})_\Theta$
will be in the
negative $\bf \hat e_\Theta$ direction: Such could very well 
balance or cancel out the effects of $(F_{_{\rm GM}})_\Theta>0$
acting on the initial infalling incident photons.
Note that, the only  PPP electrons
satisfying the escape conditions are those with $(P_\mp)_r>0$;
electrons with $(P_\mp)_r<0$ fall into the black hole since
the PPP ($\gamma \gamma \longrightarrow e^- e^+$\/) 
takes place at the photon orbit
[5].    
Now, on the other hand, at larger energies, as in the cases of 
Figs.~8(b)$-$ 8(d) and~9(b)$-$9(d), 
the initialized asymmetry, produced by $(F_{_{\rm GM}})_\Theta$ 
acting on $(P_{\gamma 1})_r<0$ (of the infalling incident 
photon), 
becomes less and less balanced as $L_\mp$ increases (implying 
increasing $E_\mp$), as  the condition 
$\vert H^r L_\mp\vert >\vert \tilde H^\Phi (P_\mp)_r\vert$
becomes  satisfied, 
at which $(F_{_{\rm GM}})_\Theta$ behaves 
like an antisymmetric function
about the equatorial plane, with $(F_{_{\rm GM}})_\Theta<0$ for 
$\Theta_\mp\leq 90^\circ$ and  $(F_{_{\rm GM}})_\Theta>0$ for
$\Theta_\mp>90^\circ$, for $(P_\mp)_r>0$, superimposed with the 
established initial
 preference for $\Theta_\mp>90^\circ$.  
So, underlining the
cause of the symmetry and asymmetry we observe in Figs.~8 and~9,
in general,  is a tug-of-war between the dominance of the 
symmetry producing term $H^rL_\mp$ [reflecting momentum transferred
from the initial 
conditions $L_{\gamma 2}$, $(P_{\gamma 2})_\Theta$]
and the asymmetry producing term 
$\tilde H^\Phi (P_{\gamma 1})_r$; more on this below.  
Note,  like the PCS, 
the way in which the GM 
field acts directly on the particles in these 
PPP ($\gamma \gamma \longrightarrow e^- e^+$\/) processes,
wherein we do not have to apply Eq.~(\ref{eq:47}) manually
as in these analyses,
is through the presence of 
$\omega$, the frame dragging
velocity [cf.~Eqs.~(\ref{eq:24}) through~(\ref{eq:30})], related to GM
potential (see  Section~\ref{sec:2d}), contained in the Kerr metric
[Eq.~(\ref{eq:1})]: The Kerr metric defines the paths of spacetime
trajectories of particles interacting and ``flowing'' along geodesics, 
with four-momenta $P_\mu=(P_r,P_\Theta,L, -E)$, in the gravitational 
field (or curved spacetime) of a 
rotating black hole, as measured by an observer at infinity.

Looking further at  the different cases in Figs.~8 and~9,
for the  case of Figs.~8(a) and~9(a) it appears that 
$(F_{_{\rm GM}})_\Theta$
[Eq.~(\ref{eq:47})]  is somewhat balanced, as described above,
 in such a way  that the 
 final result is a 
nearly symmetric distribution---like that expected from 
a Newtonian
point of view---with the ratio 
$\epsilon_\mp\simeq 1.12$ (defining the number
of $e^-e^+$ pairs scattered below to those scattered above  
the equatorial plane).
In the cases of Figs. 8(b) to 8(d) 
the ratio $\epsilon_\mp\simeq 1.74$, $\simeq 2.51$,
$\simeq 2.61$, respectively.     
Thus, in light of the general conclusion above,
 the distributions of Figs. 8(b) to 8(d) are
 consistent with both terms in
$(F_{_{\rm GM}})_\Theta$ [Eq.~(\ref{eq:47})]  
being significant:  the first term
gives the high energy PPP electrons a 
boost into opposite polar directions [cf.~Fig.~1(a)];
the second term acting in dominance 
on the incident infalling photons after 
the PPP electrons satisfy
the condition $\vert H^r L_\mp\vert>\vert \tilde H^\Phi 
(P_\mp)_r\vert$, provides the asymmetry, favoring the  positive
${\bf \hat e_\Theta}$ direction.

Overall, in these PPP ($\gamma \gamma \longrightarrow e^- e^+$\/) 
processes, we find that the intrinsic 
``$e^-e^+$ jet'' 
(positive $\bf \hat e_\Theta$ direction)
to counter jet  (negative $\bf \hat e_\Theta$ direction)
asymmetry (defined by $\epsilon_\mp$) 
increases, first quickly, as above (cf. Fig.~8), 
then slowly, to a ratio of
$\,\simeq 3:1$, for the maximum energy that can be used
for the target photon in these processes,
$E_{\gamma 2}\sim 108$~GeV (i.e., before the computer simulation
 breaks
down; 
see [5] for 
detailed construction of computer codes).
Astrophysically, what this means is that as $E_{\gamma 2}$ 
is increased,
implying increasing the energy-momentum of the escaping PPP electrons,
the luminosities in the jet ($L_{\rm Jet}$) and counter jet
($L_{\rm CJet}$)  get more one-sided,
mimicking the effects associated with relativistic beaming,
until a  maximum ratio in luminosity ($\,\simeq~3:1$) is reached.
We see this occurring when we let these PPP electrons undergo 
``secondary'' PCS
with infalling
disk photons [25], allowing us to obtain $\gamma$-ray luminosities.   
It is presumed that such a $\,\simeq~3:1$
ratio in jet luminosities would
also exist in synchrotron radiation if the PPP electrons were
allowed to interact with a surrounding magnetic field.  
Thus, it appears
that relativistic beaming near line of sight of the observer, in
general, will be needed
to explain observations if the observed $L_{\rm Jet}/L_{\rm CJet}$, 
 in a 
source powered
by a black hole, is greater than
$\sim 3$ for  the high energy $\gamma$-ray or synchrotron emission.
Nevertheless, importantly, these Penrose processes acting in 
conjunction with the GM field can serve the purpose of
beaming the escaping particles into the polar direction,
probably creating the initial jets of  AGNs.
The intrinsic magnetic field produced by the dynamo-like 
action associated with the
swirling (vortical orbiting [26]) escaping Penrose 
produced highly relativistic  plasma, about the polar axis, 
may be important also, at least in the observed synchrotron emission
of the jets, and perhaps in maintaining collimation:  
this statement must
be investigated further.

In addition, when analyzing the GM radial 
component $(F_{_{\rm GM}})_r$ of 
Eq.~(\ref{eq:47}), using Eqs.~(\ref{eq:49}), ~(\ref{eq:40}), 
and  Fig.~2(a), 
to see what effects this component has on the
PPP electrons,  we get the following. 
For $e^-e^+$ pairs scattered with $(P_\mp)_\Theta\leq 0$,  
$(F_{_{\rm GM}})_r>0$ [cf. Fig.~2(a)].  But
for $(P_\mp)_\Theta>0$,
$(F_{_{\rm GM}})_r>0$ for 
 $\vert H^\Phi(P_\mp)_\Theta\vert<
\vert H^\Theta (P_\mp)_\Phi\vert$---as would be for 
the higher energy particles; otherwise
$(F_{_{\rm GM}})_r<0$---as would be for the lower energy
particles or particles with large $(P_\mp)_\Theta>0$.  
Since in this case,  only the PPP electrons
with $(P_\mp)_r>0$ escape (on the average one of the pairs
escapes, while the other falls into the black hole [5]), 
it is not clear what role the radial component of the GM
force plays.
However, looking back at the radial spectra of the PCS photons,
this behavior appears to be a contributing factor
to the progressive asymmetry of 
$(P_{\rm ph}^\prime)_r$ seen in the escaping photons with
large $(P_{\rm ph}^\prime)_\Theta>0$ after being scattered by 
equatorially confined target electrons [cf. Figs.~3(a)$-$3(e) 
and~4(a)$-$4(e)]; and the reestablishing the symmetry seen in 
$(P_{\rm ph}^\prime)_r$ for the high energy particles 
[implying large $(P_{\rm ph}^\prime)_\Phi$] after being scattered 
by the nonequatorially confined target electrons [cf. Figs.~3(f)$-$3(i)
and~4(f)$-$4(i)].

So, overall, for the PCS and the PPP 
($\gamma \gamma \longrightarrow e^- e^+$\/) considered here,
in general, the asymmetries seen in the scattered
particle  distributions, above and below the equatorial plane, 
that one would, in nonrotating  space 
(or pseudo flat space), expect to be 
symmetrical,
 can be attributed to the GM field acting on the incident and
scattered particles.  Specifically, it is  a tug-of-war between
conditions $\vert H^r P_\Phi\vert{<\atop>}\vert \tilde H^\Phi 
P_r\vert$ defined by Eq.~(\ref{eq:47}) that causes the asymmetry,
which varies through a ratio of $\sim 3:1$ in the $e^-e^+$ jet 
to counter jet for  PPP ($\gamma \gamma \longrightarrow e^- e^+$\/), 
and a ratio of
$\sim 5:1$ in the photon jet to counter jet for PCS.
Thus,  the  
GM force, more or less, breaks  the expected reflection  
symmetry, above 
and below the
equatorial plane.   These Penrose processes, which  occur near 
the event horizon, 
allow us to see
this general relativistic effect in action. 


%
\section{Conclusions}
\label{sec:4}

It has been shown in this investigation that 
Penrose scattering processes in the
ergosphere of a KBH can produce x-ray, $\gamma$-ray, and $e^-e^+$
pair energies
comparable to those observed in quasars and other AGNs.  This
model can work for any size accretion disk (thin or ``thick'') and 
any size black hole. 
It has been shown also that the asymmetry in the distribution of
Penrose  scattered particles,
above and
below the equatorial plane, can be attributed to the GM 
field---the gravitational 
force field produced by the angular 
momentum of the KBH, which causes local
inertial frames to be dragged into the 
direction that the black hole is rotating
(sometimes called the Lense-Thirring effect). 
It has become apparent through these Penrose processes that
the GM force field lines, inside the ergosphere, 
close to the event horizon, are frame dragged
  into the positive azimuthal direction
 in the local frame of the observer
at infinity, resulting in particle distributions  appearing to 
break the expected reflection symmetry of the Kerr metric
(which describes the
geometry outside of a  rotating black hole) above and below the 
equatorial plane. 

Overall, these model calculations strongly suggest that
the momenta of the Penrose scattered particles naturally 
aid in the production of one-sided and two-sided polar jets,
consistent with the jets associated with AGNs. 
Moreover, the  variations between symmetry and asymmetry 
found in  the scattered particle 
distributions, and thus in densities, due to the GM field acting
on the space momenta of the incoming and outgoing
particles, could contribute 
to the overall individual ``blob-like''
 structures we observe in the jets 
of AGNs;  this should,  however, be investigated further.

Importantly, we have the following 
conclusion from this investigation:
The astrophysical conceptual fact
 is that rapidly rotating black holes give rise to 
Penrose processes and strong GM forces, which in turn 
produce high energy
particle jets with symmetrical and 
asymmetrical distributions: differing
by a factor as high as  $\sim 3-5$,  
above and below the equatorial plane.  This must be taking
into account in AGN models when applying 
relativistic beaming effects
[27,28,29,30].

In addition, this investigation suggests an answer
to the long asked question concerning whether 
or not the asymmetry in
the jet and counter jet distributions is 
intrinsic to the energy source [31].
The suggested 
answer, in general, is  yes, although Doppler boosting due to
relativistic 
beaming [32,33] when the emission axis
of the jet is near the line
of sight of the observer,
 may still
in some cases  
 serve as an 
important enhancement mechanism, to increase the
apparent observed luminosities and energies, particularly when
superluminal motion in the jet flow is apparent and/or the jet
and counter jet luminosities differ by more than a factor  
$\sim 3-5$.  Details of how these
Penrose scattering processes relate 
to AGNs, their observed luminosities, and relativistic beaming
models, are discussed in other papers by the author [25,34].  

Further, the model presented here to extract energy-momentum from 
a rotating black hole by Penrose escaping particles, along with
the intrinsic magnetic field associated with this relativistic 
vortical orbiting escaping ``plasma,'' which is expected to 
give rise to
synchrotron radiation---being consistent with observations,
suggests a complete, self-consistent model, without the
necessity of the external magnetic field of the accretion disk.
The attempted use of the external magnetic field of the 
accretion disk in a dominant role to extract rotational energy
from a black hole, by having the magnetic field lines torque
 the horizon and/or the accretion material near the horizon
(i.e., the Blandford-Znajek (BZ) type models; see [35] and 
references
therein), introduces historical
problems yet to be solved, namely, how to convert from 
electromagnetic
energy to relativistic particle energy, such that particles
are ejected along field lines into the polar direction;
and related problems, such as how to keep the field lines
from becoming tangled (or to be sufficiently tangled); 
how to generate the large magnetic field
strengths needed (1) if the relativistic particles are to be 
created along the
field lines and (2) to be consistent with the most luminous
AGNs.   Equally the causality problem as related to the ``vacuum
infinity'' horizon at which upon nearing, the magnetic flux is
redshifted away poses a problem for models
assuming the magnetic field lines are ``anchored'' to the
horizon or surrounding ergospheric region; see [36].
However, 
the electromagnetic field of the accretion disk could, perhaps,
play an important role on a large scale in assisting the 
jets of
the Penrose escaping particles, in further collimation and 
acceleration,
out to the observed kiloparsec distances of AGN jets.
 The large scale magnetic field of the accretion disk 
may also
be important in transporting angular momentum outward
enabling material to be accreted into the black hole
[37].
Now, by the same token, since the GM field provides a way through these
Penrose processes to transport angular momentum from particles 
inside the ergosphere,  outward to infinity, it too may 
also provide a 
way to transport
angular momentum outward, in the ``static'' [7] accretion disk regime
($r>2M$, i.e., where the severe frame dragging is less):
This statement seems worthy of an investigation.

In addition, the Penrose process distributions 
described here, of 
escaping particles, are consistent with the disk returning
photons [38], gravitationally focused after being emitted 
from  radii
$<r_{\rm ms}$ ($\simeq 1.2M$), the marginally stable orbit.
As measured by an observer at infinity, the PCS photons are  
reemitted with the highest energy photons
scattered along geodesics concentrated in the equatorial plane.
Notice,  cases similar to those of Figs.~3(b), 6(b) and 3(a), 6(a), 
7(a), 7(b) 
can be used to explain the 
observations of the classical stellar/galactic black-hole candidate
 Cygnus X-1 [39] and  the broad Fe K$\alpha$ emission
line at $\sim 6$~keV of the 
bright Seyfert~1 galaxy (AGN)
MCG---6-30-15 [35], respectively.   The case similar to that  
  of Figs.~3(b) and 6(b) which is consistent with Cyg X-1 has model 
parameters for radial infalling photons ($E_{\rm ph}=3.5$~keV) 
from a thin
disk [40], that either undergo PCS by  equatorially confined orbiting 
target electrons ($E_e\simeq 0.539$~MeV) or 
PPP ($\gamma \gamma \longrightarrow e^- e^+$\/) at $r_{\rm mb}$
or $r_{\rm ph}$, respectively: 
the blueshifted 
 energies (due to frame dragging) attained by 
the $\sim 82\%$ up to $90\%$ escaping particles, returning to the 
disk to be reprocessed and/or escaping
to infinity, are the following:  for the PCS photons, 
$E_{\rm ph}^\prime\sim 12-250$~keV (i.e., up to $\sim 70\, 
E_{\rm ph}$), with relative incoming and outgoing
photon luminosities 
$(L_\gamma)_{\rm out}\sim 0.4- 130 \,(L_\gamma)_{\rm in}$, 
respectively; 
and for the relativistic PPP
electrons (with $E_{\gamma 2}\sim 5$~MeV), 
$E\mp\sim 4$~MeV [consistent with synchrotron radiation 
in the radio regime for $B\sim 10^{2}$~gauss, and inverse Compton
scattering  (of disk photons) into the hard 
x-rays/soft $\gamma$-ray regime---with relative incoming and 
outgoing photon luminosities  $(L_\gamma)_{\rm out}\sim 8-2000 \,
(L_\gamma)_{\rm in}$, for $M\sim 30M_\odot$, at 
$\sim 100$~keV$-3$~MeV, respectively]---[cf. also 
Figs.~8(a) and 9(a)];
see [5,25,34,41] for details.   
The case similar to those of Figs.~3(a), 6(a), 7(a), 7(b), 
which is consistent with MCG---6-30-15,  has model
parameters for  radial infalling photons ($E_{\rm ph}=1.5$~keV)
from a thin
disk, that either undergo PCS by  
equatorially or nonequatorially
confined orbiting
target electrons ($E_e\simeq 0.539$~MeV or 
$E_e\simeq 0.615$~MeV, respectively) at $r_{\rm mb}$ or
PPP ($\gamma \gamma \longrightarrow e^- e^+$\/) at $r_{\rm ph}$: 
the blueshifted
 energies (due to frame dragging) attained by
the $\sim 72\%$ up to $95\%$ escaping particles, returning to the
disk to be reprocessed and/or escaping
to infinity, are the following:  for PCS photons,
$E_{\rm ph}^\prime\sim 4.6-138$~keV 
or $E_{\rm ph}^\prime\sim 6-169 $~keV  
for equatorially or nonequatorially confined orbiting
target electrons, respectively, 
with relative photon luminosities
$(L_\gamma)_{\rm out}\sim 0.03-14\,(L_\gamma)_{\rm in}$
or $(L_\gamma)_{\rm out}\sim 0.04-27\,(L_\gamma)_{\rm in}$,
respectively;
and for the relativistic PPP
electrons (with $E_{\gamma 2}\simeq 6.8$~MeV), 
$E\mp\sim 3.5$~MeV [consistent with synchrotron radiation
in the radio regime for $B\sim 10^{2}$~gauss, and inverse Compton
scattering (of disk photons) into the 
 hard
x-ray/soft $\gamma$-ray  regime---with relative incoming and
outgoing photon luminosities $(L_\gamma)_{\rm out}\sim 0.07-20 \,
(L_\gamma)_{\rm in}$, for $M\sim 10^8M_\odot$, 
at $\sim 103~{\rm keV}-2$~MeV, respectively], 
suggesting  relatively weak, less powerful 
and less prominent radio
jets, i.e., a radio quiet AGN, like a Seyfert galaxy; see [5,25,34,41] 
for details.  Note, $E_{\gamma 2}$ is assumed based on prior PCS
photons with $(P_{\rm ph}^\prime)_r<0$ that satisfy conditions 
for the existence of a turning point at the photon orbit 
(discussed in Section~\ref{sec:2c}), and on satisfaction of the energy 
threshold for the particles to react (see [5]).
Thus, the consistency of
the results of these Penrose process analyses with observations 
strongly suggests that
the appropriate way to exact energy from a KBH is by  the 
``Penrose-Williams'' mechanism, described here and in 
[5,25,34,41]: using gravity as the controlling force---to distinguish it 
from BZ-type models, which use electromagnetism as the controlling force.

Lastly, I want to conclude with the comment, that, 
the physics described in
this paper and the associated paper of [5], can be applied to
 different types of high energy-momentum 
exchange scattering processes, i.e.,
processes that would be expected to occur in the ``hot'' ergosphere
of a rotating black hole, the intention being to extract rotational 
energy-momentum. An immediate future investigation would be to
perform the proton-proton nuclear scattering of neutral pions
$\pi^\circ$, and their subsequent 
$\pi^\circ\longrightarrow\gamma\gamma$ decay
(see [5,8,20]),  inside the ergosphere of a KBH.  These 
$\gamma$-rays
are presumed to be important in the population of the photon
orbit in the highest energy 
PPP ($\gamma \gamma \longrightarrow e^- e^+$\/) 
processes.  Note, similar nuclear scattering and $\pi^\circ$-decay 
processes, however, for particles confined to the 
equatorial plane are investigated in [3].
%
%

%
%
\acknowledgments
First, I thank God for His thoughts and for making this 
research possible.
Next, I wish to thank Dr.~Henry Kandrup for helpful comments 
and discussions,
in particular those
concerning the gravitomagnetic field, 
and for his careful reading of this 
manuscript.  I am grateful to Drs.~Fernando de Felice and 
Bernard Whiting for their comments and 
suggestions. Also, I thank Dr.~Roger Penrose for his continual 
encouragement.  Finally, I thank North Carolina A \& T State University 
for their
hospitality while this work was being completed.  Part of this work
was done at the Aspen Center for Physics.  This work was
supported in part by the National Science Foundation and Bennett
College.
%

%

%
\vfill\eject
%
%
%
\vglue 2in
\begin{figure}[tbh] 
\centerline{\epsfig{file=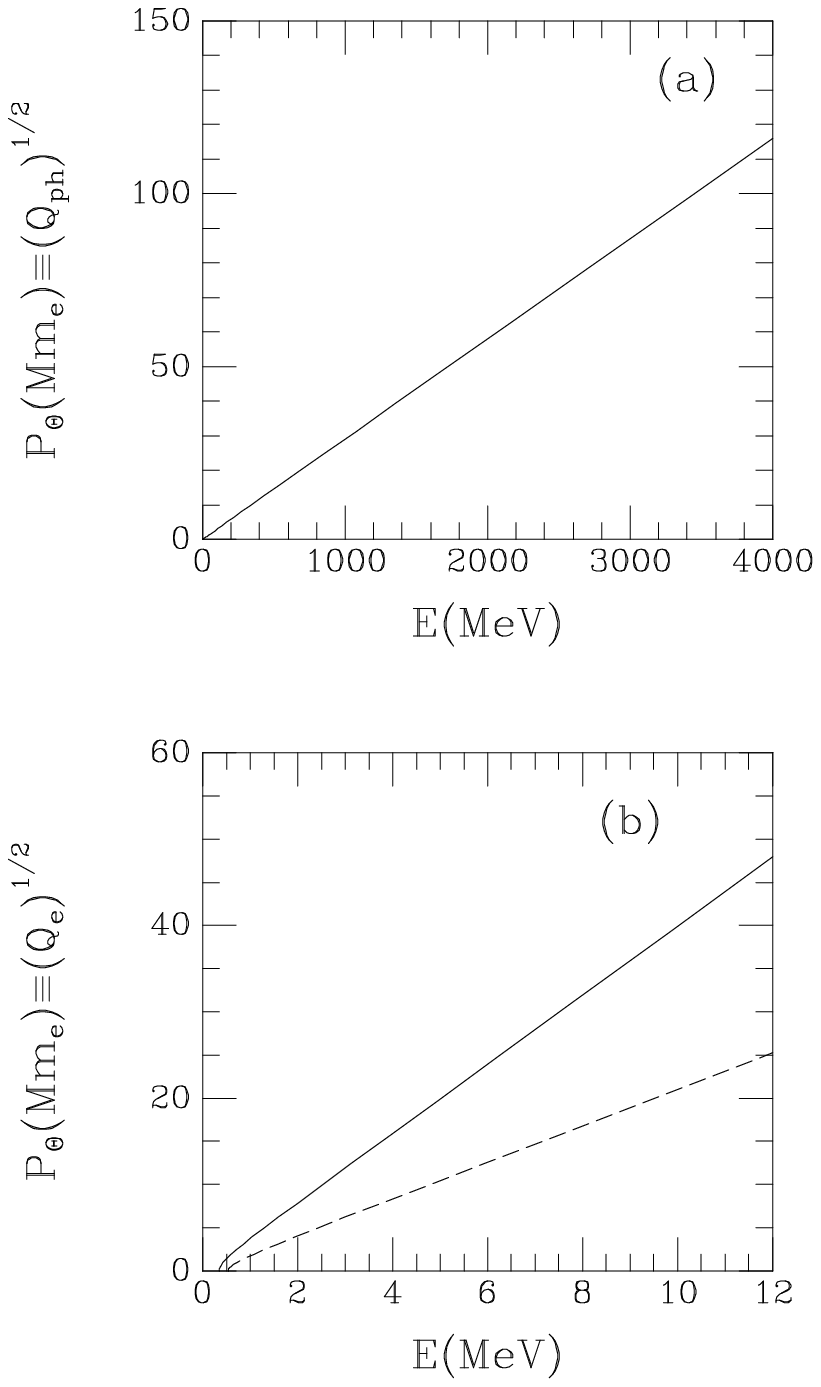,height=2.in,width=2.in,
bbllx=2.2in,bblly=7.0in,bburx=4.0in,bbury=9.0in,angle=0.
}}
\vskip 2.9in
\caption{Orbital parameters of massless and material particles.
Magnitude of polar coordinate angular momentum $P_\Theta$
as the bound orbit crosses the equatorial plane ($\equiv Q^{1/2}$)
vs. the conserved nonequatorially confined orbital energy:
(a) $(P_{\rm ph})_\Theta$
vs. $E_{\rm ph}$ at the photon orbit, $r_{\rm ph}\simeq 1.074M$.
(b) $(P_{e})_\Theta$ vs. $E_e$, of an orbiting electron, at
$r_{\rm mb}\simeq 1.089M$ (dashed curve) and  at
$r_{\rm ms}\simeq 1.2M$ (solid curve).
Notice that, as these momenta
go to zero, each of the orbital energies goes to its equatorial
confinement value---i.e., as $Q_{\rm ph}^{1/2}\longrightarrow 0$,
$E_{\rm ph}\longrightarrow 0$;
as $Q_e^{1/2}\longrightarrow 0$,
$E_e\longrightarrow 0.539$ MeV at $r_{\rm mb}$,
and $E_e\longrightarrow 0.349$ MeV
at $r_{\rm ms}$. (Note that, when the more
exact value of $r_{\rm mb}=1.091M$
is used, $E_e\longrightarrow 0.512$~MeV $\simeq \mu_e$.)}
\label{figone}
\end{figure}
\vfill\eject
                                                                                \vglue 3in
\begin{figure}[tbh] 
\centerline{\epsfig{file=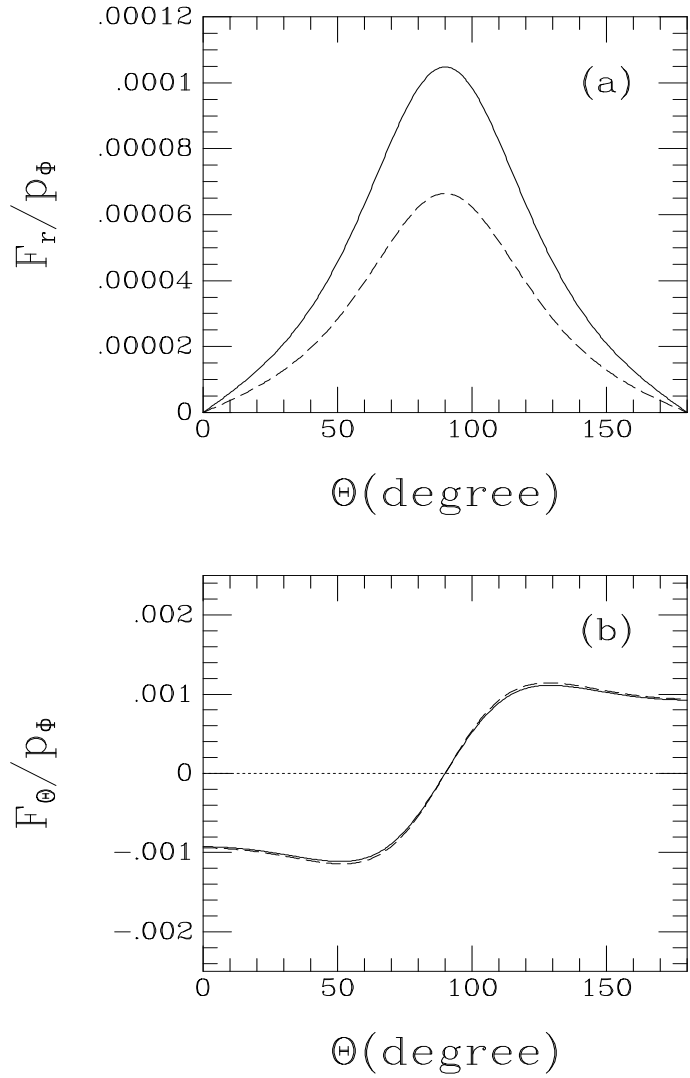,height=2.in,width=2.in,
bbllx=2.2in,bblly=6.0in,bburx=4.0in,bbury=8.0in,angle=0.
}}
\vskip 2.in
\caption{
The GM field force components.  (a) The radial component:
$(f_{_{\rm GM}})
_r/p_\Phi$ vs. $\Theta$. (b) The polar coordinate component:
$(f_{_{\rm GM}})_\Theta/p_\Phi$ vs. $\Theta$.
The solid curves indicate the force
at $r_{\rm mb}\simeq 1.089M$; the dashed
curves indicate the force at
$r_{\rm ph}\simeq 1.074M$.}
\label{figtwo}
\end{figure}
\vfill\eject
                                                                                \vglue 3in
\begin{figure}[tbh] 
\centerline{\epsfig{file=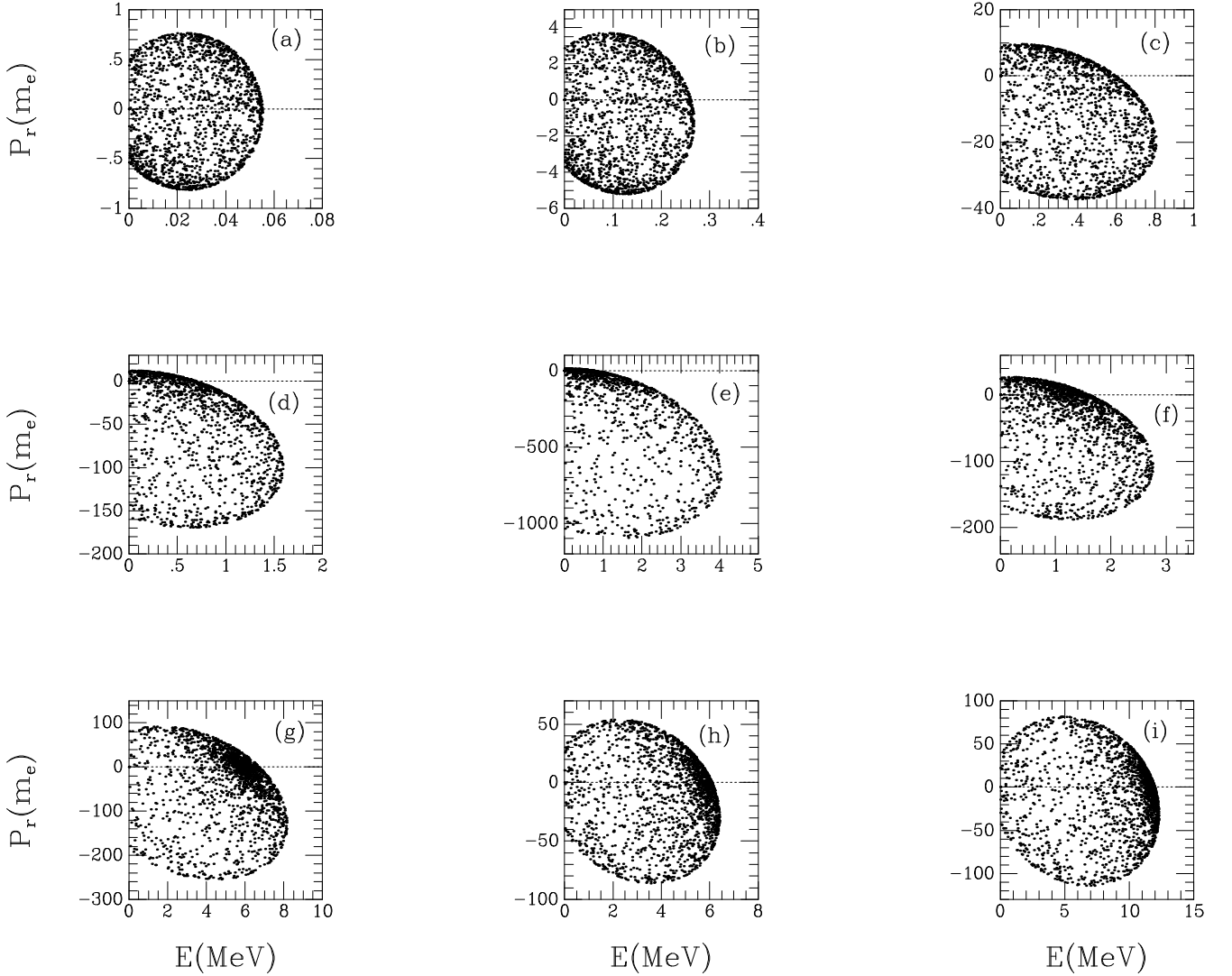,height=2.in,width=2.in,
bbllx=3.1in,bblly=6.0in,bburx=5.15in,bbury=8.0in,angle=0.
}}
\vskip 2.0in
\caption{
Compton~scattering: scatter plots
showing radial momentum components:
$(P_{\rm ph}^\prime)_r$ vs. $E_{\rm ph}^\prime$,
of scattered escaping photons after 2000 events
(each point represents a
scattering event), at $r_{\rm mb}\simeq 1.089M$.
The various cases are defined
by the following
parameters: $E_{\rm ph}$, initial photon energy;
$E_e$,  target electron orbital energy;
$Q_{\rm e}^{1/2}$, corresponding polar coordinate
momentum $(P_{e})_\Theta$ of the target electron;
$N_{\rm es}$, number
of photons escaping. (a) $E_{\rm ph}=0.511$~keV,
 $E_e=0.539$~MeV, $Q_{\rm e}^{1/2}=0$,
$N_{\rm es}=1707$.
(b) $E_{\rm ph}=3.5$~keV,
 $E_e=0.539$~MeV, $Q_{\rm e}^{1/2}=0$,
$N_{\rm es}=1637$.
(c) $E_{\rm ph}=0.03$~MeV,
 $E_e=0.539$~MeV, $Q_{\rm e}^{1/2}=0$,
$N_{\rm es}=1521$.
(d) $E_{\rm ph}=0.15$~MeV,  $E_e=0.539$~MeV, $Q_{\rm e}^{1/2}=0$,
$N_{\rm es}=1442$.
(e) $E_{\rm ph}=1$~MeV,  $E_e=0.539$~MeV, $Q_{\rm e}^{1/2}=0$,
$N_{\rm es}=1390$.
(f) $E_{\rm ph}=0.15$~MeV,  $E_e=1.297$~MeV,
$Q_{\rm e}^{1/2}=\pm 2.479\,Mm_e$,
$N_{\rm es}=1628$.
(g) $E_{\rm ph}=0.15$~MeV,  $E_e=5.927$~MeV,
$Q_{\rm e}^{1/2}=\pm 12.43\,Mm_e$,
$N_{\rm es}=1843$.
(h) $E_{\rm ph}=0.03$~MeV,  $E_e=5.927$~MeV,
$Q_{\rm e}^{1/2}=\pm 12.43\,Mm_e$,                                              $N_{\rm es}=1935$.
(i) $E_{\rm ph}=0.03$~MeV,
 $E_e=11.79$~MeV, $Q_{\rm e}^{1/2}=\pm 24.79\,Mm_e$,
$N_{\rm es}=1971$. (Note, due to a minor
 oversight leading to improper treatment in the computer simulation
of the arccosine term in
 eq.~(3.39) of
[5], correct Fig.~3(i) presented here
replaces Fig.~4(a) of [5].)}
\label{figthree}
\end{figure}
\vfill\eject
 
                                                                                \vglue 3in
\begin{figure}[tbh] 
\centerline{\epsfig{file=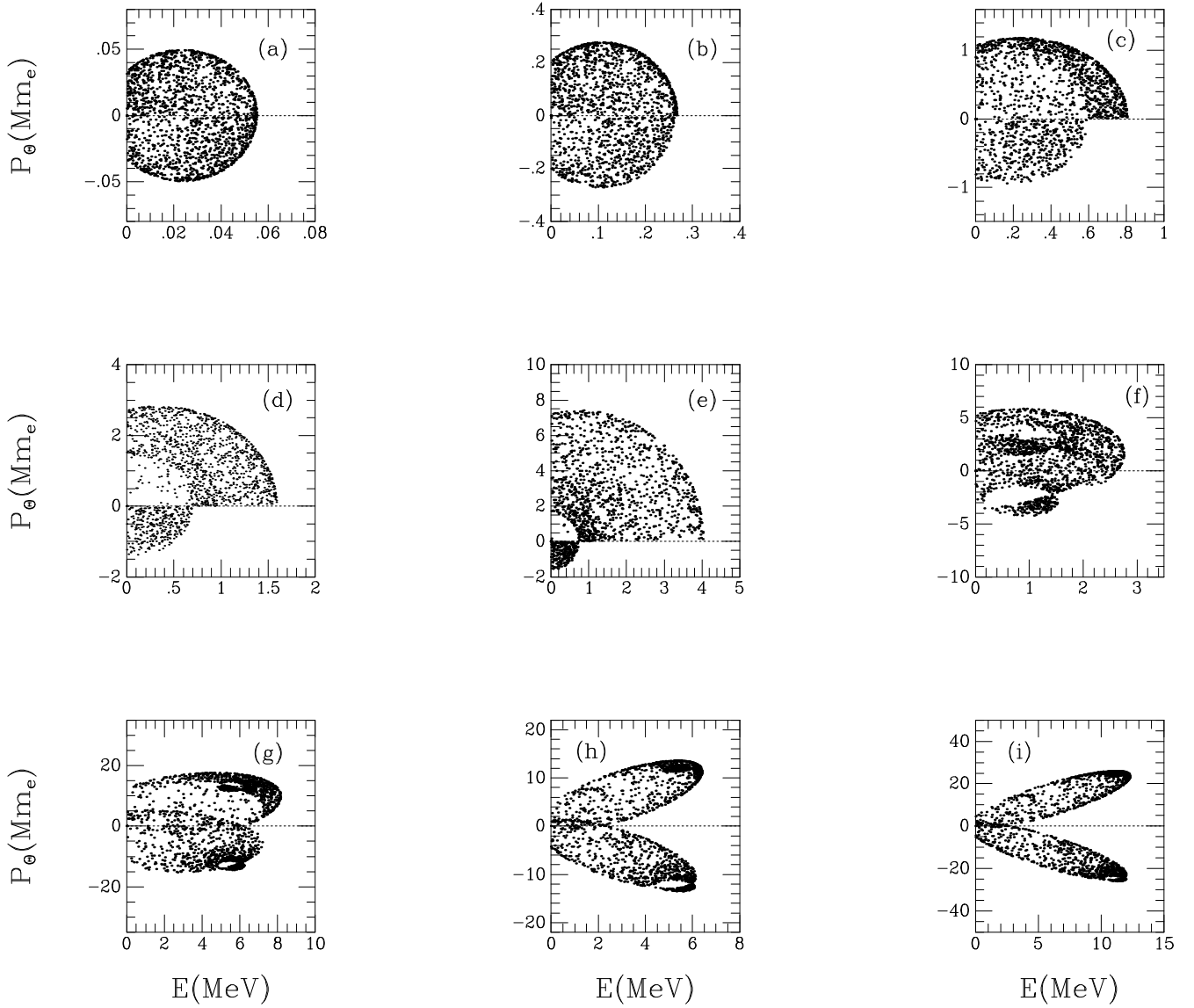,height=2.in,width=2.in,
bbllx=3.12in,bblly=6.0in,bburx=5.17in,bbury=8.0in,angle=0.
}}
\vskip 2.33in
\caption{
Compton~scattering: scatter plots showing the corresponding
polar coordinate space momentum components:
$(P_{\rm ph}^\prime)_\Theta$
$[\equiv (Q_{\rm ph}^\prime)^{1/2}$]
vs.~$E_{\rm ph}^\prime$, of the escaping PCS photons,
for the cases (a)$-$(i),
respectively, described in Fig.~3.
The units of $(P_{\rm ph}^\prime)_\Theta $ are $\,Mm_e$
($G=c=1$). (Note, due to a minor
 oversight leading to improper treatment in the computer simulation
of the arccosine term in
 eq.~(3.39) of
[5], correct Figs.~4(b) and~4(c) presented here
replace Figs.~7(a)
and~3(c), respectively, of [5].)}
\label{figfour}
\end{figure}
\vfill\eject
                                                                                \vglue 3in
\begin{figure}[tbh] 
\centerline{\epsfig{file=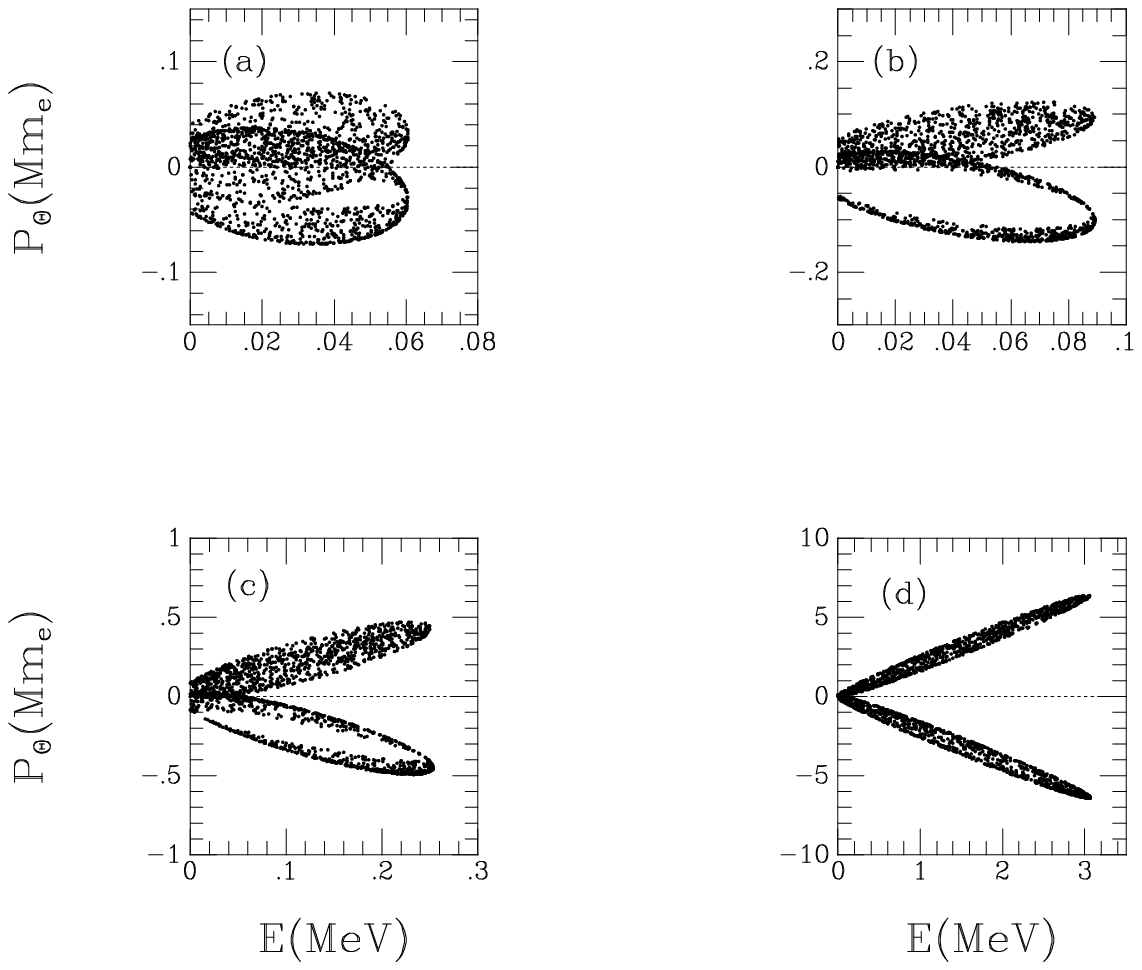,height=2.in,width=2.in,
bbllx=2.2in,bblly=6.0in,bburx=4.0in,bbury=8.0in,angle=0.
}}
\vskip .26in
\caption{Compton scattering: scatter plots showing polar
coordinate space momentum components:
$(P_{\rm ph}^\prime)_\Theta$
$[\equiv (Q_{\rm ph}^\prime)^{1/2}$]
vs.~$E_{\rm ph}^\prime$, of the escaping PCS photons after
2000 events
(each point represents a
scattering event), at $r_{\rm mb}\simeq 1.089M$.
The various cases are defined
by the following
parameters: $E_{\rm ph}$, initial photon energy;
$E_e$,  target electron orbital energy;
$Q_{\rm e}^{1/2}$, corresponding polar coordinate
momentum $(P_{e})_\Theta$ of the target electron;
$N_{\rm es}$, number
of photons escaping.
(a) $E_{\rm ph}=0.511$~keV,  $E_e=0.570$~MeV,
$Q_{\rm e}^{1/2}=\pm 0.393\,Mm_e$,
$N_{\rm es}=1719$.
(b) $E_{\rm ph}=0.511$~keV,  $E_e=0.714$~MeV,
$Q_{\rm e}^{1/2}=\pm 0.987\,Mm_e$,
$N_{\rm es}=1779$.
(c) $E_{\rm ph}=0.511$~keV,  $E_e=1.297$~MeV,
$Q_{\rm e}^{1/2}=\pm 2.479\,Mm_e$,
$N_{\rm es}=1896$.
(d) $E_{\rm ph}=0.511$~keV,  $E_e=5.927$~MeV,
$Q_{\rm e}^{1/2}=\pm 12.43\,Mm_e$,
$N_{\rm es}=1996$.
}
\label{figfive}
\end{figure}
\vfill\eject
                                                                                \vglue 3in
\begin{figure}[tbh] 
\centerline{\epsfig{file=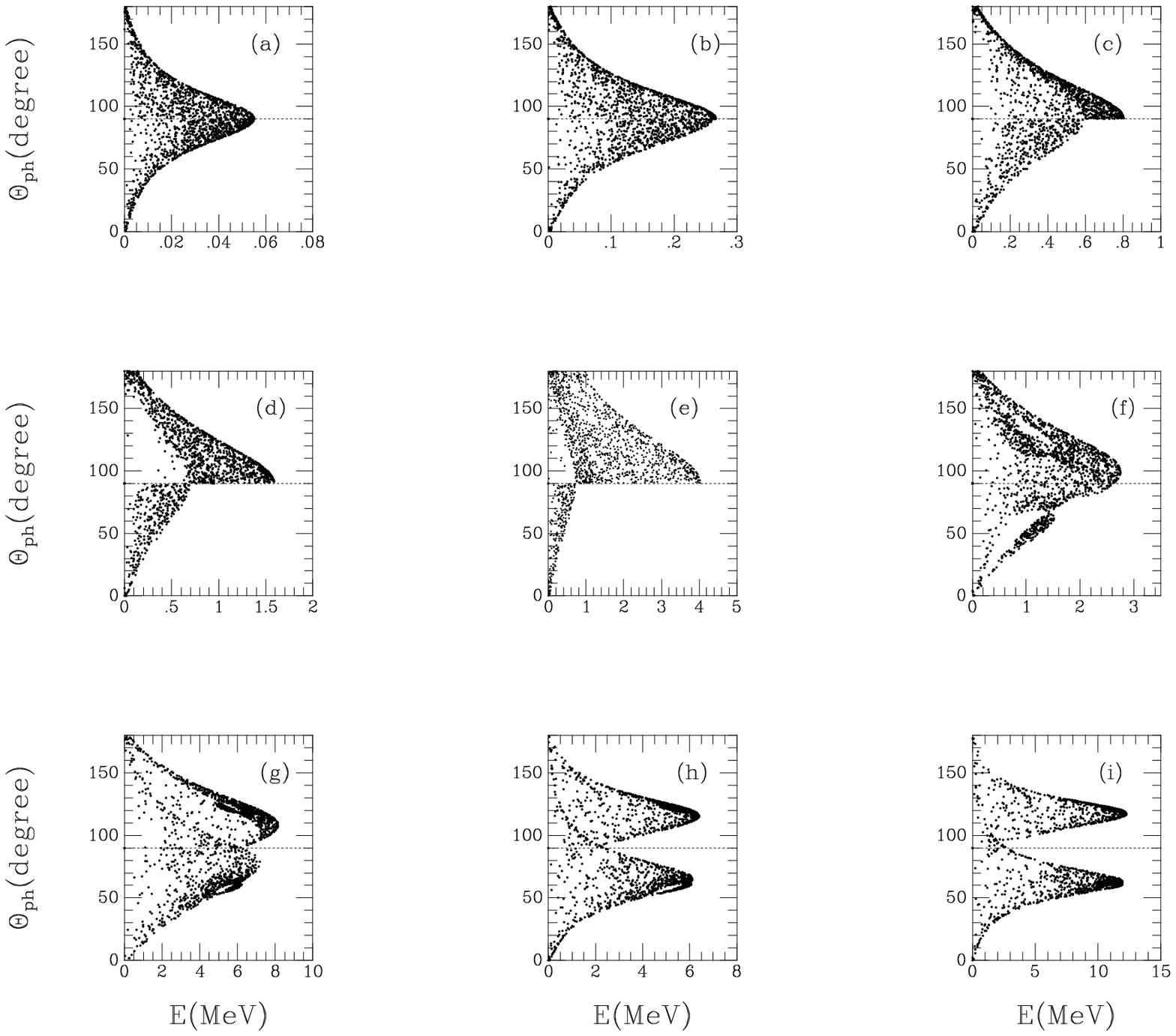,height=2.in,width=2.in,
bbllx=3.12in,bblly=6.0in,bburx=5.17in,bbury=8.0in,angle=0.
}}
\vskip 2.7in
\caption{
Compton scattering: scatter plots
displaying polar angles, above and below the equatorial plane:
$\Theta_{\rm ph}^\prime$ vs.~$E_{\rm ph}^\prime$, of the escaping
PCS photons,
for the cases (a)$-$(i),
respectively, described in Figs.~3 and~4.
Note that, $\Theta_{\rm ph}^\prime > \pi/2$
is below the equatorial plane.}
\label{figsix}
\end{figure}
\vfill\eject
                                                                                \vglue 3in
\begin{figure}[tbh] 
\centerline{\epsfig{file=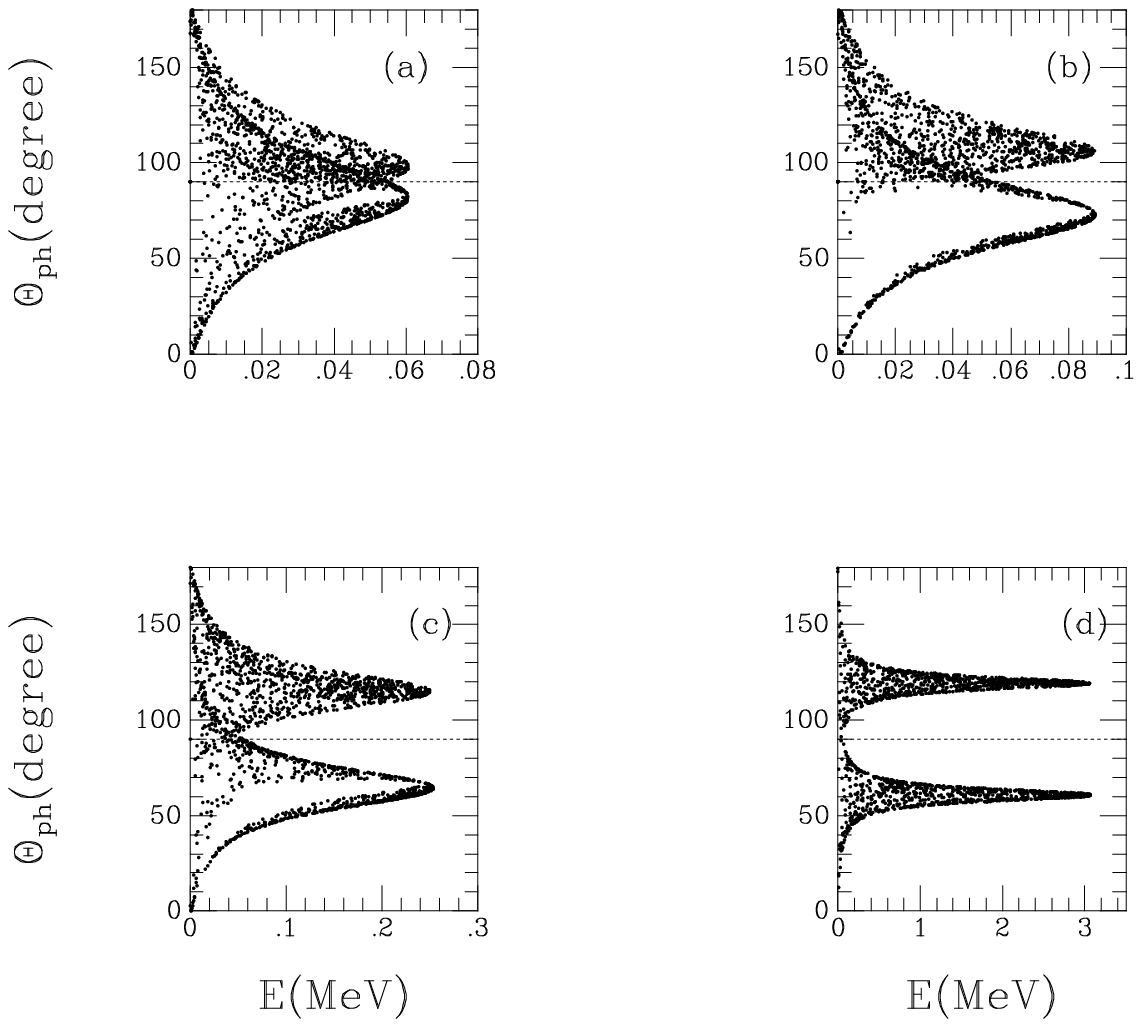,height=2.in,width=2.in,
bbllx=2.2in,bblly=6.0in,bburx=4.0in,bbury=8.0in,angle=0.
}}
\vskip 0.46in
\caption{Compton scattering: scatter plots
displaying polar angles, above and below the equatorial plane:
$\Theta_{\rm ph}^\prime$ vs.~$E_{\rm ph}^\prime$, of the escaping
PCS photons,
for the cases (a)$-$(d),
respectively, described in Fig.~5.
Note that, $\Theta_{\rm ph}^\prime > \pi/2$
is below the equatorial plane.}
\label{figseven}
\end{figure}
\vfill\eject
                                                                                \vglue 3in
\begin{figure}[tbh] 
\centerline{\epsfig{file=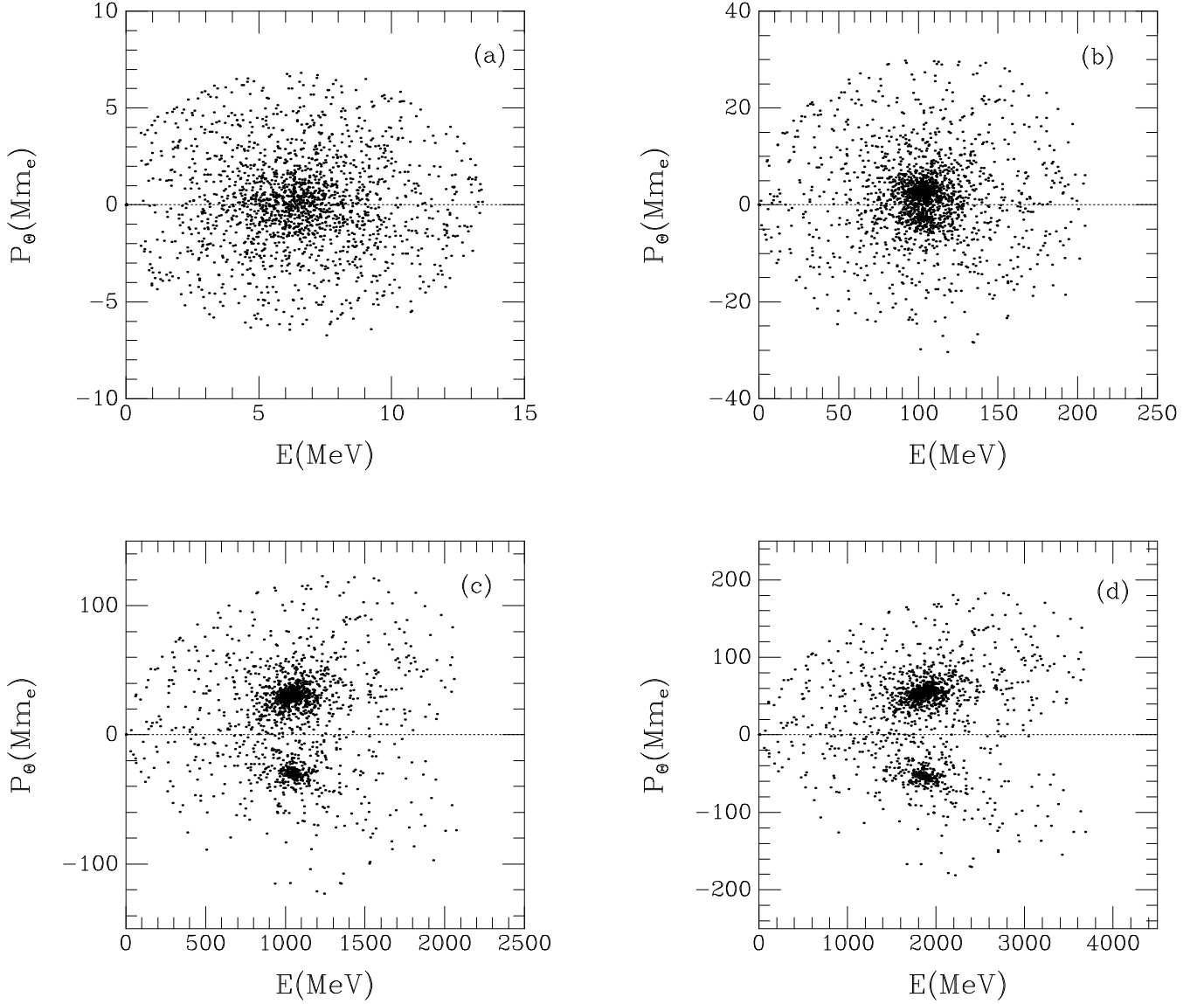,height=2.in,width=2.in,
bbllx=3.11in,bblly=6.0in,bburx=5.16in,bbury=8.0in,angle=0.
}}
\vskip 2.35in
\caption{
Penrose~pair~production ($\gamma \gamma \longrightarrow e^- e^+$\/):
scatter plots showing
polar coordinate space momentum components:
$P_\Theta $ ($\equiv Q^{1/2}$)
vs. $E_\mp $,
of scattered escaping $e^-e^+$ pairs after 2000 events
(each point represents a
scattering event), at $r_{\rm ph}$.
The various cases are defined
by the following
parameters: $E_{\gamma 1}$, the infalling photon energy;
$E_{\gamma 2}$, the target photon orbital energy;
$Q_{\gamma 2}^{1/2}$, corresponding polar coordinate
momentum $(P_{\gamma 2})_\Theta$ of the target photon;
$N_{\rm es}$, number
of  $e^-e^+$ pairs escaping. (a) $E_{\gamma 1}=0.03$~MeV,
 $E_{\gamma 2}=13.54$~MeV,
$Q_{\gamma 2}^{1/2}=\pm 0.393\,Mm_e$,
$N_{\rm es}=1850$.
(b) $E_{\gamma 1}=0.03$~MeV,  $E_{\gamma 2}=206.7$~MeV,
$Q_{\gamma 2}^{1/2}=\pm 6.0\,Mm_e$,
$N_{\rm es}=1984$.
(c) $E_{\gamma 1}=0.03$~MeV,  $E_{\gamma 2}=2.146$~GeV,
$Q_{\gamma 2}^{1/2}=\pm 62.28\,Mm_e$,
$N_{\rm es}=1997$.
(d) $E_{\gamma 1}=0.03$~MeV,  $E_{\gamma 2}=3.893$~GeV,
$Q_{\gamma 2}^{1/2}=\pm 113.0\,Mm_e$,
$N_{\rm es}=1997$.}
\label{figeight}
\end{figure}
\vfill\eject
\vglue 3in
\begin{figure}[tbh] 
\centerline{\epsfig{file=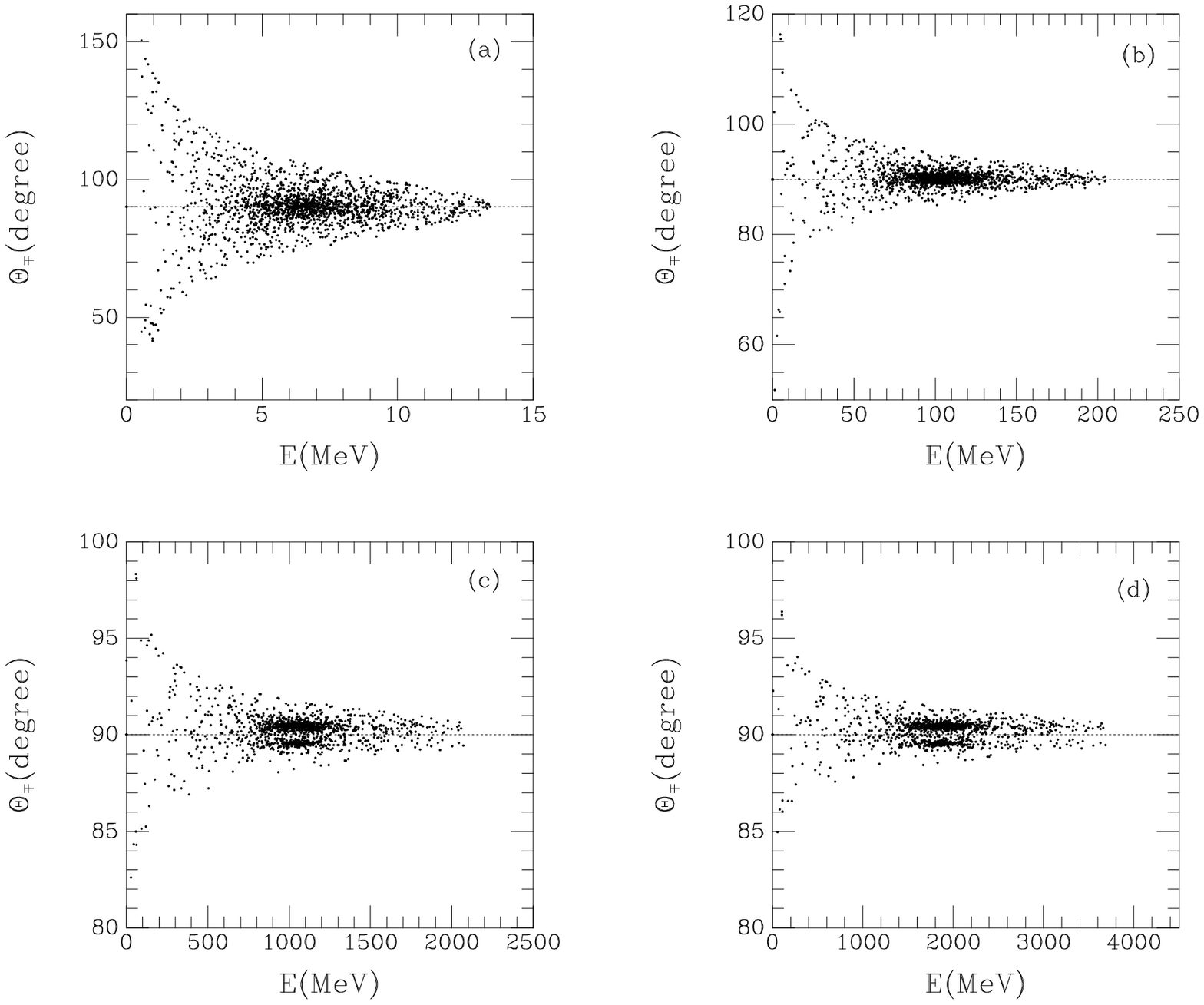,height=2.in,width=2.in,
bbllx=3.13in,bblly=6.0in,bburx=5.18in,bbury=8.0in,angle=0.
}}
\vskip 2.3in
\caption{
Penrose pair production ($\gamma \gamma \longrightarrow e^- e^+$\/):
scatter plots
displaying polar angles, above and below the equatorial plane:
$\Theta_\mp$ vs.~$E_\mp^\prime$, of the escaping
$e^-e^+$ pairs,
for the cases (a)$-$(d), respectively,
described in Fig.~8.}
\label{fignine}
\end{figure}

\begin{references}
\bibitem{key1}H. Thirring and J. Lense, Phys. Z. $\bf 19$, 156 (1918). 
\bibitem{key2}R. Penrose, Rivista Del Nuovo Cimento: Numero
  Speciale $\bf 1$, 252 (1969).
\bibitem{key3}T. Piran and J. Shaham, Phys. Rev. D. $\bf 16$, No. 6,
  1615 (1977).
\bibitem{key4}T. Piran and J. Shaham, Astrophys. J.
   $\bf 214$, 268 (1977).
\bibitem{key5}R. K. Williams, Phys. Rev. D $\bf 51$, 5387 (1995).
\bibitem{key6}T. Piran, J. Shaham, and J. Katz, Astrophys. J. Lett.
  $\bf 196$, L107 (1975).
\bibitem{key7}J. M. Bardeen, W. H. Press, and S. A. Teukolsky,
  Astrophys. J. $\bf 178$, 347 (1972).
\bibitem{key8}J. A. Eilek and M. Kafatos, Astrophys. J. $\bf 271$,
  804 (1983).
\bibitem{key9}D. Leiter and M. Kafatos, Astrophys. J. $\bf 226$,
  32 (1978); M. Kafatos and D. Leiter, Astrophys. J. $\bf 229$,
  46 (1979).
\bibitem{key10}K. S. Thorne, R. H. Price, and D. A. Macdonald,
{\it Black Holes: The Membrane Paradigm}
(Yale University Press,
New Haven, 1986).
\bibitem{key11}D. L. Jones, in {\it Superluminal
Radio Sources},
edited by J. A. Zensus and T. J. Pearson
(Cambrige University Press,
Cambrige, 1987).
\bibitem{key12}J. Dennett-Thorpe, A. H. Bridle, P. A. G. Scheuer,
R. A. Laing, and J. P. Leahy, Mon. Not. R. Astron. Soc. $\bf 289$,
753 (1997).
\bibitem{key13}I. F. Mirabel and L. F. Rodriguez, Nature $\bf 392$,
673 (1998).
\bibitem{key14}R. M. Hjellming and M. P. Ruben, Nature $\bf 375$,
464 (1995).
\bibitem{key15}R. P. Kerr, Phys. Rev. Letters $\bf 11$, 237 (1963). 
\bibitem{key16}R. H. Boyer and R. W. Lindquist, J. Math. Phys. $\bf 8$,
   265 (1967).
\bibitem{key17}K. S. Thorne, Astrophys. J.  $\bf 191$, 507 (1974).
\bibitem{key18}B. Carter, Phys. Rev. $\bf 174$, No. 5, 1559 (1968).
\bibitem{key19}D. C. Wilkins, Phys. Rev. D {\bf 5}, 814 (1972).
\bibitem{key20}J. A. Eilek, Astrophys. J. $\bf 236$, 664 (1980).
\bibitem{key21}S. L. Shapiro, A. P. Lightman, and D. M. Eardley,
   Astrophys. J. $\bf 204$, 187 (1976).
\bibitem{key22}A. P. Lightman, Astrophys. J. {\bf 194}, 429 (1974).
\bibitem{key23}R. Mahadevan,  R. Narayan, and J. Krolik,
Astrophys. J. $\bf 486$, 
268 (1997).
\bibitem{key24}J. A. Eilek (private communication).
\bibitem{key25}R. K.  Williams, submitted to Astrophys. J.,
astro-ph/0306135.
\bibitem{key26}F. de Felice and O. Zanotti,
Gen. Relativ. Gravit. $\bf 32$, 1449 (2000); F. de Felice and L.
Carlotto, Astrophys. J.  $\bf 481$, 116 (1997).
\bibitem{key27}R. D. Blandford, C. F. McKee, and M. J. Rees,
 Nature $\bf 267$, 211 (1977).
\bibitem{key28} R. D. Blandford and A. K{\"o}nigl,
 Astrophys. J. $\bf 232$, 34 (1979).
\bibitem{key29}C. D. Dermer, R. Schlickeiser,
 and A. Mastichiadis, Astron. Astrophys. $\bf 256$, L27 (1992).
\bibitem{key30}L. Maraschi, G. Ghisellini, G., and 
A. Celotti, Astrophys. J. Lett.
  $\bf 397$,   L5 (1992).
\bibitem{key31}P. D. Barthel, in {\it Superluminal
Radio Sources},
edited by J. A. Zensus and T. J. Pearson
(Cambrige University Press,
Cambrige, 1987).
\bibitem{key32}A. H. Bridle and R. A. Perley,
Ann. Rev. Astr. Ap. $\bf 22$,
319 (1984).
\bibitem{key33}P. Parma, C. Fanti, R. Fanti,
  R. Morganti, and H. R. de Ruiter,
    Astron. Astrophys. $\bf 181$, 244 (1987).
\bibitem{key34}R. K. Williams, in {\it Relativistic 
Astrophysics: 20th Texas Symposium},
edited by J. C. Wheeler and H. Martel
(American Institute of Physics,
New York, 2001), astro-ph/0111161.
\bibitem{key35}J. Wilms $\it et al$., Mon. Not. R. Astron. Soc., 
$\bf 328$, L27 (2001), astro-ph/0110520.
\bibitem{key36}B. Punsly and F. V. Coroniti, Phys. Rev. D,
$\bf 40$, 3834 (1989); Astrophys. J. $\bf 350$,
518 (1990); $\bf 354$,
583 (1990); B. Punsly, Astrophys. J. $\bf 372$, 424 (1991);
B. Punsly and D. Bini, Astrophys. J. $\bf 601$, L135 (2004);
J. Bi\v{c}\'{a}k, Pramana $\bf 55$, No. 4, 481 (2000), gr-qc/0101091;
J. Bi\v{c}\'{a}k and T. Ledvinka, IL Nuovo Cimento $\bf 115~B$, 739
(2000), gr-qc/0012006.
\bibitem{key37}S. A. Balbus and J. F. Hawley, Astrophys. J.
$\bf 376$, 214 (1991).
\bibitem{key38}C. T. Cunningham, Astrophys. J. $\bf 202$, 788 (1975);
C. Cunningham, Astrophys. J. $\bf 208$, 534 (1976).
\bibitem{key39}E. P. Liang, Phys. Rep. $\bf 302$, 67 (1998).
\bibitem{key40}I. D. Novikov and K. S. Thorne, in {\it Black Holes},
edited by C. DeWitt and  B. S. DeWitt (Gordon and Breach, New York,
  1973). 
\bibitem{key41}R. K. Williams, submitted to Astrophys. J.,
astro-ph/0210139.
 \end{references}
\end{document}